\documentclass[superscriptaddress,prd,showpacs,twocolumn,showpacs,nofootinbib, toc=flat]{revtex4}

\usepackage{float}
\usepackage{graphicx}
\usepackage{color}
\usepackage{epsf}
\usepackage{bm}
\usepackage{amsmath}
\usepackage{amsfonts}
\usepackage{amssymb}
\usepackage{epstopdf}%
\newcommand{\lsim}   {\mathrel{\mathop{\kern 0pt \rlap
  {\raise.2ex\hbox{$<$}}}
  \lower.9ex\hbox{\kern-.190em $\sim$}}}
\newcommand{\gsim}   {\mathrel{\mathop{\kern 0pt \rlap
  {\raise.2ex\hbox{$>$}}}
  \lower.9ex\hbox{\kern-.190em $\sim$}}}
\newcommand{\bw}{\begin{widetext}\begin{equation}}
\newcommand{\ew}{\end{equation}\end{widetext}}
\newcommand{\be}{\begin{equation}}
\newcommand{\ee}{\end{equation}}
\newcommand{\bea}{\begin{eqnarray}}
\newcommand{\eea}{\end{eqnarray}}

\begin{document}

\title{Testing  Bose-Einstein Condensate dark matter models with the SPARC galactic rotation curves data}

\author{Maria Cr${\rm \breve{a}}$ciun}
\email{craciun@ictp.acad.ro}
\affiliation{Institute of Numerical Analysis, Cluj-Napoca, Romania}
\author{Tiberiu Harko}
\email{tiberiu.harko@aira.astro.ro}
\affiliation{Astronomical Observatory, 19 Ciresilor Street, 400487 Cluj-Napoca, Romania,}
\affiliation{Department of Physics, Babes-Bolyai University, 1 Kogalniceanu Street,
400084 Cluj-Napoca, Romania}
\affiliation{School of Physics, Sun Yat-Sen University,  Xingang  Road, 510275 Guangzhou, People's
Republic of China}

\date{\today}

\begin{abstract}
The nature of one of the fundamental components of the Universe, the dark matter, is still unknown. One interesting possibility is that dark matter could exist in the form of a self-interacting Bose-Einstein Condensate (BEC). The fundamental properties of the dark matter in this model are determined by two parameters only, the mass and the scattering length of the particle.  In the present study we investigate the properties of the galactic rotation curves in the BEC dark matter model, with quadratic self-interaction,  by using 173 galaxies from the recently published Spitzer Photomery \& Accurate Rotation Curves (SPARC) data. We fit the theoretical predictions of the rotation curves in the slowly rotating BEC models with the SPARC data by using genetic algorithms. We provide an extensive set of figures of the rotation curves,  and we obtain estimates of the relevant astrophysical parameters of the BEC dark matter halos (central density, angular velocity and static radius). The density profiles of the dark matter distribution are also obtained. It turns out that the BEC model gives a good description of the SPARC data. The presence of the condensate dark matter could also provide a solution for the core/cusp problem.
\end{abstract}


\maketitle
\tableofcontents

\section{Introduction}

 The possible presence at different length scales in the Universe of different forms of yet unseen and undetected  matter is one of the basic postulates of the present day dominant cosmological paradigm, known as the $\Lambda$ Cold Dark Matter ($\Lambda$CDM) model. Its success essentially relies on the assumption of the existence of two dark components in the Universe, dark energy, and dark matter, respectively. The reality of dark matter, first suggested a long time ago in \cite{zw1} and \cite{zw2},  is required by the necessity of explaining two fundamental astrophysical observations. The first one is the behavior of the galactic rotation curves.  In spiral and many other galaxy types one can observe neutral hydrogen clouds at large distances
from the galactic center, and much outside the limits of the presence of the luminous matter. The clouds
rotate in circular orbits around the galactic center with velocity $v_{tg}(r)$. The stable dynamical orbits are maintained by the balance between the Newtonian gravitational force $GM(r)/r^2$ generated by the total mass $M(r)$ located within the orbit, and
the centrifugal acceleration $v^2_{tg}/r$, which gives for the tangential velocity the standard Keplerian expression $v_{tg}=\sqrt{GM(r)/r}$. Hence, according to this simple Newtonian result, outside the baryonic matter distribution, and at large distances from the galactic center, the tangential velocity should decay to zero. However, many astrophysical observations
do  show that the rotational velocities first increase near the center of the galaxy,
and then remain nearly constant at a value of the order of $v_{tg\infty} \approx 100-200$ km/s \cite{RC1,RC2,RC3,RC4,RC5,RC6}. Note that this implies a mass
profile of the form $M(r) = rv^2_{tg\infty}/G$, and indicates that the mass within a distance $r$ from the galactic center increases linearly with the distance, even at large distances $r$ where very little baryonic matter can be detected observationally.

The second important astrophysical information about the existence of dark matter comes from the study of the clusters
of galaxies. There are two ways that allow  the estimation of the total mass of a cluster. First of all, after determining the kinetic energies
of the member galaxies, with the use of the virial theorem, one can give an estimate of the mass $M_V$ of the cluster. A second estimate is
obtained by independently computing the mass of each individual galaxy, and then summing up all these
masses. This procedure gives the total baryonic mass $M_B$. Almost without exception it is found that $M_V$ is
much larger than $M_B$, $M_V > M_B$, with the typical values of the ratio $M_V /M_B$ taking values of the order of  20-30 \cite{RC1}.  The consistent  explanation of other  types of observations, like, for example, gravitational lensing, also require the introduction of dark matter \cite{l1,l2,l3}. Dark matter is also necessary to elucidate the chemical composition of the Universe, as it results from  the measurements of the cosmological
parameters using the properties of the  Cosmic Microwave Background Radiation. The recent data obtained by the Planck satellite have shown that the Universe
is composed of 4\% baryons, 22\% non-baryonic dark matter and 74\% dark energy \cite{Planck}.

The nature and the physical properties of dark matter and of its physical constituents (the dark matter particles) have been investigated extensively from many points of view (for reviews and in depth discussions on the role of dark matter in astrophysics and cosmology see \cite{rev1,rev2,rev3,rev4,rev5}. Popular candidates for the dark matter particle  are the Weakly Interacting Massive Particles (WIMP), but no significant detections of them have been reported yet \cite{wimp1, wimp2}. Present day observations of the projected surface stellar density rule out at the 99.9\% confidence level the possibility that more than 6\% of the dark matter is composed of black holes with a mass of few tens of solar masses \cite{loeb1}.  By using the large velocity dispersion of Eri II combined with the existence of a central star cluster constraints on the number of massive compact halo object dark matter with mass $M\geq 10 M_{\odot}$ have been obtained in \cite{loeb2}.

An alternative possibility for the explanation of the current observations is based on the assumption that at galactic or extra-galactic scales Einstein's general relativity (including its Newtonian limit) breaks down, and new theories of gravity are needed for the understanding of the galactic, extra-galactic and cosmological dynamics, with  the laws of gravity drastically revised at large scales.
Modified theories of gravity have been used extensively as  alternative explanations to dark matter \cite{m1,m2,m3,m4,m5,m6,m7,m8,m9,m10,m11,m12}.

One of the most important and substantial direct evidences for the existence of dark matter are provided by the weak lensing observations of a galaxy cluster known as the Bullet Cluster \cite{bc1,bc2}. The Bullet Cluster is the high velocity (around 4500 km/s) collision of two clusters, with the gases in the clusters interacting as the merger develops. The gas in the clusters slows down, and heats up and emits X-rays. But the gas lags behind the subcluster galaxies. On the other hand,  the dark matter clump, which can be reconstructed from the weak lensing map, is synchronous with the collisionless galaxies, but is situated ahead of the collisional gas.

Even very successful in providing good qualitative explanations for the behavior and constancy of the rotation curves, and of the dynamics of galaxy clusters, dark matter models face a number of serious challenges, yet to be answered.  One major problem arises from the confrontation of the results of the numerical simulations performed in the framework of the $\Lambda $CDM model, and observations.  The numerical simulations imply
a central dark matter density profile behaving according to
$\rho  \sim 1/r$ (thus forming a cusp) \cite{cusp}. On the other hand the observed rotation curves show the presence of a significant constant density cores \cite{core1,core2}. Hence it turns out that the observational data on rotation curves indicate a less abrupt increase as compared with the predictions of the cosmological
simulations of structure formation in the standard
$\Lambda CDM$ model, based on the fundamental hypothesis of a single pressureless dark matter fluid. In the physics and astrophysics of dark matter this major contradiction represents the so-called core-cusp problem.

A second important open question dark matter models need to address is the too big to fail problem \cite{tbf1,tbf2,tbf3}.
The Aquarius simulations performed within the framework of the $\Lambda$CDM model indicate that the predicted number of the most massive subhalos in the dark matter halos is in strong conflict with the observational detections of the  dynamics of the brightest dwarf spheroidal galaxies of the Milky Way  \cite{tbf2}. Many important observations show that the best-fitting hosts of the dwarf spheroidals galaxies have velocities of the order of  $12 < V_{max} < 25$ km/s. On the other hand all, $\Lambda$CDM numerical simulations predict a number of at least ten subhalos with $V_{max} > 25$ km/s. These results cannot be understood in the framework of the $\Lambda$CDM models of the satellite population of the Milky Way. Here the key problem, necessary to be addressed by all dark matter models, is the density of the satellites. This problem does appear due to the fact that in the simulations the dwarf spheroidals must have dark matter halos five times more massive than indicated by the astronomical observations.

These important and intriguing problems standard dark matter theories must face could be easily solved by assuming that dark matter is self-interacting \cite{si1,si2,si3,si4,si5}. This hypotheses may be supported observationally by the study of the data obtained from the observations of 72 cluster collisions, including both major as well as minor mergers,  performed by the Chandra and Hubble Space Telescopes \cite{Bul, Bul1, Bul2, Bul3}. Hence important tests of the non-gravitational forces acting on
dark matter can be obtained from the analysis of the collisions between galaxy clusters.
The effect of dark matter self-interaction on the tidal stripping and evaporation of satellite galaxies in a Milky Way-like hosts was investigated in \cite{si8}.  Self-interacting dark matter  models provide excellent fits to the rotation curves for galaxies  with asymptotic velocities in the range of 25 to 300 km/s \cite{si9}. Note that in dark matter dominated galaxies, thermalization due to self-interactions creates large cores, and reduces dark matter densities. However, in more luminous galaxies  thermalization leads to denser and smaller cores, and thus naturally explains the flat rotation curves of the highly luminous galaxies. The implications of the self-interacting dark matter halos for the understanding of the growth of galaxy potentials, using idealized numerical simulations, were considered in \cite{si10}.

To analyze the response of self-interacting dark matter  halos in the presence of the baryonic potentials N-body simulations were performed in \cite{si11}. It was thus shown that dark matter self-interactions can lead to the kinematic thermalization in the inner halo, hence giving rise to a tight correlation between the dark matter and baryon distributions. The final density profile is sensitive to the baryonic concentration and the strength of the dark matter self-interactions.  Numerical simulations were used in \cite{si12} to analyze the effects of dark matter interactions on the morphology of disk galaxies falling into galaxy clusters. The effective drag force on dark matter causes distortions in the disk, thus leading to a balance of the stellar disk with respect to the surrounding halo.

However, from a physical point of view the most reasonable candidate for self-interacting dark matter is the Bose-Einstein Condensate dark matter model. Bose-Einstein Condensation in bosonic systems is a well known and much studied phenomenon, both from experimental and theoretical point of view. Initially introduced by Bose \cite{Bose}, and further and deeply developed by Einstein  \cite{Ein1,Ein2}, the study of the Bose-Einstein Condensation has become a fundamental field of study, especially after its detection in laboratory experiments \cite{exp1,exp2,exp3}. The Bose-Einstein Condensation process occurs when the temperature $T$ of the boson gas drops below the critical temperature $T_{cr}$, given by \cite{re1,re2,re3,re4}
\begin{equation}  \label{Ttr}
T_{cr}=\frac{2\pi\hbar^2\rho_{cr}^{2/3}}{ \zeta^{2/3}(3/2)m^{5/3}k_B},
\end{equation}
where $m$ is the particle mass in the condensate, $\rho_{cr}$ is the
critical transition density, $k_B$ is Boltzmann's constant, and $\zeta $ denotes the
Riemmann zeta function. Note that from a physical point of view the Bose-Einstein Condensation process develops due to the quantum mechanical correlation of the boson gas particles, which happens when the de Broglie thermal wavelength turns out to be greater than the mean interparticle distance. A well acknowledged  phenomenon in the laboratory, Bose-Einstein Condensation may also materialize in a cosmic environment. A significant amount of matter inside high density neutron stars can exist in the form of a Bose-Einstein Condensate \cite{starsm1, stars0, stars1,stars2,stars3,stars4,stars5,stars6,stars7, stars8}. Stars consisting or containing  a Bose-Einstein Condensate may have maximum masses of the order of 2 $M_{\odot}$, maximum central densities in the range $0.1-0.3\times 10^{16}$ g/cm$^3$, and minimum radii of the order of 10-20 km, respectively.

Since dark matter is assumed to be formed from a bosonic gas, and if we postulate that the laws of physics in the Universe are the same as on Earth, the possibility that a Bose-Einstein Condensation process may have arisen during the cosmological evolution must be taken very seriously into account. This idea  was first  proposed  in \cite{early1}, and then reinvestigated or rediscovered in \cite{early2, early3,early4,early5, early6,early7, early8, early9a, silverman2002dark, rotha2002vortices, early9}. The systematic and in depth investigation of the properties of the Bose-Einstein Condensate (BEC) dark matter halos, based on the hydrodynamic formulation of the non-relativistic Gross-Pitaevskii (GP) equation in the presence of a confining gravitational potential,  was initiated in
\cite{BoHa07}.  The theoretical investigations of the gravitationally bounded BECs can be significantly simplified by introducing the Madelung (hydrodynamic)
representation of the wave function. This permits to write down the GP
equation in a form similar to classical fluid mechanics, thus enhancing both the range of its applications, as well as its physical interpretation.  That is, the GP equation reduces in fact to a continuity equation, and to a hydrodynamic Euler type equation in the presence of a quantum pressure, and a quantum potential, respectively. By using the hydrodynamic representation for the dark matter BEC confined by the gravitational trapping potential one can arrive at the fundamental result that dark matter can be described as a non-relativistic gas,  with the pressure and density obeying a standard polytropic type equation of state, with polytropic index $n=1$. The investigation of the properties of the Bose - Einstein
Condensate dark matter models, as well as of their cosmological and astrophysical implications, is presently a very active and important field of research \cite{HaM,Har1,inv0, inv1, inv2,inv3,inv4,inv5,inv6,inv7,inv8,inv9,inv10,inv11,inv12,inv13,inv14,inv15,inv16,inv17,inv18,inv19,inv20,inv21,inv22,inv23,inv24,inv25,inv26, inv27,inv28,inv29,inv30,inv31,inv32,inv33,inv34,inv35, inv37,Chavn, Hui,Chav1, Mem3,Zhang, inv36, inv36a, inv38,inv39, inv40,inv41, inv42}.

Recently a sample of 175 nearby galaxies with new surface photometry at 3.6 $\mu$m and high-quality rotation curves from previous HI/Halpha studies has been published. The corresponding database is called SPARC (Spitzer Photometry \& Accurate Rotation Curves) \cite{Lelli1,Lelli2}. SPARC spans a broad range of morphologies (S0 to Irr), luminosities (~5 dex), and surface brightnesses (~4 dex). Detailed mass models have also been built, and the ratio of baryonic-to-observed velocity ($V_{bar}/V_{obs}$) for different characteristic radii and values of the stellar mass-to-light ratio (M/L) at [3.6] was quantified.

The SPARC data have been employed in \cite{Lelli3} to study correlation between the radial acceleration traced by rotation curves and that predicted
by the observed distribution of baryons.   The same database has been used in \cite{Bern} to study the rotation curves with two Scalar Field Dark Matter  profiles, the soliton+Navarro-Frenk-White profile in the fuzzy dark matter model, arising empirically from cosmological simulations of a real, non-interacting scalar field at zero temperature, and for the multistate Scalar Field Dark Matter profile.
The possibility whether the extra force term generated in the Cubic Galileon Gravity can explain the missing mass problem in galaxies without the help of dark matter was investigated,  by using the Milky Way rotation curve and the SPARC data in \cite{Chan}.  In \cite{Li1} the Markov Chain Monte Carlo method was used to fit the mean tight radial acceleration relation to the 175 individual galaxies in the SPARC database.  For the vast majority of galaxies acceptable fits with astrophysically reasonable parameters were  found, with the residuals  having an rms scatter of only 0.057 dex ($\sim 13$\%). In \cite{LMond} it was attempted to differentiate between modified gravity and modified inertia with galaxy rotation curves from the SPARC sample. Rotation curve fits to 175 late-type galaxies from the SPARC, database using seven dark matter halo profiles: pseudo-isothermal, Burkert, Navarro-Frenk-White, Einasto, Di Cintio, coreNFW, and Lucky13 were presented in \cite{Gough20}.

The correlations between the intrinsic properties of galaxies
and the Bose-Einstein Condensate  dark matter were investigated, by assuming that the temperature of the condensate is greater than zero, in \cite{New}. By using the SPARC dataset, it was found that the condensate dark matter parameters present a weak correlation only
with most of the galaxy properties. A correlation does exist only with the galactic properties
related to neutral hydrogen emissions. In this study evidence for the
self-interaction between the different BEC states was also found, and it was shown that there is a null correlation
with galaxy distances.

The central prediction of the BEC dark matter model, according to which  the equation of state of the dark matter is the  $n=1$ polytropic one, leads to a Lane-Emden type nonlinear second order differential equation for the density profile.  The solution of this equation can be obtained in an exact analytical form, in terms of elementary functions. Note that the density (as well as the pressure) vanishes at the vacuum boundary of the dark matter halo. This important condition fixes the value of the static (nonrotating) halo radius as a simple function of the two fundamental physical parameters fully describing the properties of the condensed dark matter, namely, the mass $m$ of the dark matter particle, and its scattering length $a$, respectively.

The effects of the rotation on the structure of condensate dark matter halos were investigated in \cite{Zhang}.  By assuming a rigid body rotation for the halo, with the use of the hydrodynamic representation of the Gross-Pitaevskii equation, the basic equation describing the density distribution of the rotating condensate was obtained. The general solutions for the condensed dark matter density was obtained, by also taking into account the effects of the rotation, and the general expressions for the mass distribution, radius and several other astrophysical parameters were obtained. The results were compared with the observations by fitting the theoretical expressions of the tangential velocity of massive test particles with the data of 12 dwarf galaxies, and the Milky Way, respectively.

It is the main goal of the present paper to investigate the high quality SPARC galactic rotation curves sample \cite{Lelli1,Lelli2} in the framework of the Bose-Einstein Condensate dark matter model. In this model one can obtain explicit analytical representations of the rotation velocity of massive particles moving on circular trajectories in condensate galactic halos, as well as of the mass and density distributions of the halo.
 After performing several cuts to the SPARC sample, we restrict our analysis to a number of 85 galaxies, with the baryonic matter component modeled by a gaseous disc, a stellar
disc, and a stellar bulge (if appropriate), respectively. For the dark matter halo we adopt the description of the standard Bose-Einstein Condensate dark matter model, by assuming that it consists of a rotating $n=1$ polytrope.  The theoretical model is fitted with the observations  by adopting  a fitting technique based on genetic algorithms, and the physical parameters for each galaxy (central density, rotational velocity, and halo radius) are obtained. Moreover, we provide an extensive set of figures of the rotation curves. Overall, we find that the Bose-Einstein Condensate dark matter model gives a good description of the SPARC sample, with around 75\% of the performed fittings having a value of $\chi ^2$ smaller than 3.

The present paper is organized as follows. We briefly present the Bose-Einstein Condensate dark matter model in Section~\ref{sect1}, where the expressions of the rotational velocity and of the density distribution are obtained. The comparison of the theoretical model with the ${\rm SPARC}$ sample is performed in Section~\ref{sect2}, where we present the fitting of a large number of rotation curves, as well as the predicted dark matter halo density profiles. The astrophysical properties of the BEC dark matter halos are also obtained. We discuss and conclude our results in Section~\ref{sect3}.

\section{Dark matter as a Bose-Einstein Condensate}\label{sect1}

The starting point in the study of the Bose-Einstein Condensate dark matter in the second quantization is the Hamiltonian operator of the interacting bosons, which is generally given by \cite{re1, Nat1}
\bea
\hat{H}&=&\int d^3\left(\vec{r}\right)\hat{\Psi}^{+}\left(\vec{r}\right)\Bigg[-\frac{\hbar ^2}{2m}\Delta +U\left(\vec{r}\right)+U_{rot}\left(\vec{r}\right)+\nonumber\\
&&\frac{1}{2}\int d^3\left(\vec{r}\;'\right)U_{int}\left(\vec{r}-\vec{r}\;'\right)\hat{\Psi}^{+}\left(\vec{r}\;'\right)\hat{\Psi}\left(\vec{r}\;'\right)\Bigg]\hat{\Psi}\left(\vec{r}\right),\nonumber\\
\eea
where  $\hat{\Psi}\left(\vec{r}\right)$ and  $\hat{\Psi}^{+}\left(\vec{r}\;'\right)$  are the annihilation and creation operators at the position $\vec{r}$, $m$ is the mass of the particle in the dark matter condensate, $U$ denotes the external potential, $U_{rot}$ is the effective centrifugal potential, while $U_{int}$ is the  interparticle interaction potential. Note that in order to study the rotational properties of the condensate we will adopt the comoving frame, that is, the frame that is rotating at the same velocity as the condensed dark matter halo.

When discussing  Bose-Einstein Condensates one usually assumes that the interparticle interaction is a short range one. Therefore we can write the interaction potential as the product between a constant $u_0$ that  depends on the scattering length, and a Dirac delta function \cite{re1,re2,re3,re4},
\begin{equation}
\begin{aligned}
U_{int}\left(\vec{r}\;'-\vec{r}\right)=&u_0\delta\left(\vec{r}\;'-\vec{r}\right),
\end{aligned}
\end{equation}
where we have denoted
\be
u_0=\frac{4\pi a\hbar^2}m,
\ee
 with $a$ representing the scattering length.  After  introducing the mean field description, in which the field operator in the Heisenberg picture is given by  $\hat{\Psi}(\boldsymbol{r},t)=\psi(\boldsymbol{r},t)+\hat{\Psi}'(\boldsymbol{r},t)$, where $\psi(\boldsymbol{r},t)=\langle\hat{\Psi}(\boldsymbol{r},t)\rangle$ is  called the condensate wave function, and by integrating the Heisenberg equation of motion, we obtain the Gross-Pitaevskii equation, which can be used to effectively describe  the main properties of a Bose-Einstein Condensate, as \cite{re1,re2,re3,re4}
\bea\label{gp}
  i\hbar\frac{\partial}{\partial t}\psi(\vec{r},t)&=&\Big[-\frac{\hbar^2}{2m}\nabla^2+U\left(\vec{r}\right)+U_{rot}\left(\vec{r})\right)
  +\nonumber\\
 &&+ u_0|\psi(\vec{r},t)|^2)\Big]\psi\left(\vec{r},t\right).
\eea
The number density of the dark matter particles  in the condensate is given by $n\left(\vec{r},t\right)=|\psi\left(\vec{r},t\right)|^2$, while the normalisation condition is $N=\int {n(\boldsymbol{r},t)d^3\boldsymbol{r}}$. In the following we denote by $\rho =m n\left(\vec{r},t\right)$ the mass density of the condensed dark matter.

As for the external potential $U$, generally it can be taken as the superposition  of the mean trapping gravitational potential $\phi$ of the total amount of dark matter of the halo, of the contribution of the baryonic matter $U_b$, and of a random Gaussian potential $U_{dis}$, which represents the degree of disorder in the system, so that
\be
U\left(\vec{r}\right)=\phi \left(\vec{r}\right)+U_b \left(\vec{r}\right)+U_{dis}\left(\vec{r}\right).
\ee
 We also assume that the external potential $U $ satisfies the Poisson equation,
\be
\Delta U \left(\vec{r}\right)=4\pi Gm\left[\rho \left(\vec{r}\right)+\rho_b\left(\vec{r}\right)+\rho _{dis}\left(\vec{r}\right)\right],
\ee
where $G$ is the gravitational constant, $\rho _b$ is the baryonic matter density and $\rho _{dis}$ describes the possible random fluctuations of the matter density in the system.  Generally, one can assume that both the random potential $U_{dis}$ and the random density fluctuations are characterized by their average values \cite{Nat1, Nat3},
\be\label{dis1}
\left<U_{dis}\left(\vec{r}\right)\right>=0,\left<U_{dis}\left(\vec{r}\right)U_{dis}\left(\vec{r}\;'\right)\right>=\kappa ^2\delta \left(\vec{r}-\vec{r}\;'\right),
\ee
and
\be
\left<\rho_{dis}\left(\vec{r}\right)\right>=0,\left<\rho_{dis}\left(\vec{r}\right)\rho_{dis}\left(\vec{r}\;'\right)\right>=\kappa_1 ^2\delta \left(\vec{r}-\vec{r}\;'\right),
\ee
respectively. However, in the following we will neglect any dissipative or random effects in the dark matter condensate, and therefore we will assume that $U_{dis}=0$ and $\rho_{dis}=0$, respectively.

\subsection{The hydrodynamic representation}

As we have already mentioned, the investigation of the properties of the condensate dark matter halos is very much simplified by using the hydrodynamic representation of the Gross-Pitaevskii equation. To obtain the hydrodynamic representation of the condensate dark matter halos we represent the wave function as
\be
\psi\left(\vec{r},t\right)=\sqrt{n\left(\vec{r},t\right)}e^{iS\left(\vec{r},t\right)/\hbar},
\ee
where $S\left(\vec{r},t\right)/\hbar$ is the phase of the wave function. Then the Gross-Pitaevskii equation (\ref{gp}) can be reformulated equivalently as a continuity and hydrodynamic type Euler equation as
\be
\frac{\partial n}{\partial t}+\nabla \cdot \left(n\vec{v}\right)=0,
\ee
and
\bea\label{eu1}
m\frac{d\vec{v}}{dt}&=&m\Bigg[\frac{\partial \vec{v}}{\partial t}+\left(\vec{v}\cdot \nabla\right)\vec{v}\Bigg]=-\nabla \Bigg[U\left(\vec{r}\right)+U_{rot}\left(\vec{r}\right)+\nonumber\\
&&u_0n-\frac{\hbar ^2}{2m\sqrt{n}}\Delta \sqrt{n}\Bigg],
\eea
respectively, where we have introduced the velocity of the condensate defined as $\vec{v}=\nabla S\left(\vec{r},t\right)/m$. For a static galaxy  all times derivatives vanish identically, and we can also safely neglect the macroscopic motions (the velocity) of the halo. Moreover, we neglect the last term (corresponding to the so-called quantum potential) in Eq.~(\ref{eu1}). This is the so-called Thomas-Fermi approximation. Therefore Eq.~(\ref{eu1}) becomes
\be\label{13}
 \nabla \Bigg[U\left(\vec{r}\right)+U_{rot}\left(\vec{r}\right)+
u_0n\Bigg]=0.
\ee
 For the rotational potential we adopt the expression
\be
U_{rot}=-\omega^2r^2,
\end{equation}
where $\omega$ is the angular velocity of the dark matter halo. Hence from Eq.~(\ref{13}), after applying again the $\nabla$ operator, we obtain the basic equation describing, within the adopted approximations, the equilibrium properties of the Bose-Einstein Condensate dark matter halos,
\be\label{dens}
\Delta \rho \left(\vec{r}\right)+k^2\left[\rho \left(\vec{r}\right)+\rho _b\left(\vec{r}\right)-\frac{\omega ^2}{2\pi G}\right]=0,
\ee
where we have denoted
\be
k^2=\frac{4\pi Gm^2}{u_0}=\frac{Gm^3}{a\hbar ^2}.
\ee
From a mathematical point of view Eq.~(\ref{dens}) represents a second order Helmholtz type partial differential equation.

\subsection{Radii and masses of the static dark matter Bose-Einstein Condensates}

In the case of the quadratic nonlinearity, which we consider in the present study,  the equation of state of the dark matter condensate is given by
\begin{equation}
P(\rho)=\frac{2\pi a\hbar ^2}{m^3}\rho^2.
\ee

From the above equation of state of the condensate it follows that $P\propto\rho^2$.    A general polytropic equation of state can be written as $P\propto \rho ^{1+1/n}$, where $n$ is the polytropic index. Hence we obtain the basic result that the equation of state of the Bose-Einstein Condensate dark matter is of polytropic type, with the polytropic index of the condensate being $n=1$.

By neglecting the effects of the baryonic matter, that is, by assuming $\rho \left(\vec{r}\right)>>\rho _b\left(\vec{r}\right)$, we  obtain the basic equation giving the density profile of the rotating dark matter halo as
\begin{equation}\label{Helm}
  \nabla^2\rho+k^2\left(\rho-\frac{\omega^2}{2\pi G}\right)=0.
\end{equation}

 If the dark matter could be described by a polytropic equation of state with a different polytropic index, $n\neq 1$, instead of Eq.~(\ref{Helm})  we would obtain for the description of dark matter  a general Lane-Emden equation \cite{BoHa07}, which is not linear anymore.


Under the assumption of spherical symmetry, and assuming that the dark matter halo is nonrotating, $\omega =0$,   Eq.~(\ref{Helm}) has the simple solution \cite{BoHa07}
\begin{equation}\label{rhostat}
\rho(r)=\rho_0\frac{\sin kr}{kr},
\end{equation}
where $\rho_0$ is an arbitrary integration constant. The density distribution of the dark matter is nonsingular at the center $r=0$. The radius $R$ of the static halo, an important physical quantity characterizing dark matter, can be obtained from the boundary condition $\rho(R)=0$, from which it follows that
\bea\label{rad}
\hspace{-1cm}R&=&\frac{\pi}k=\pi \sqrt{\frac{a\hbar ^2}{Gm^3}}=\nonumber\\
\hspace{-0.5cm}&&13.5\times \left(\frac{a}{10^{-17}\;{\rm cm}}\right)^{1/2}\times \left(\frac{m}{10^{-36}\;{\rm g}}\right)^{-3/2}\;{\rm kpc}.
\eea

 For the central density we find $\rho_c=\rho(0)=\rho_0$, a relation that fixes the value of the integration constant $\rho _0$. From the expression of the radius of the dark matter distribution, given by Eq.~(\ref{rad}), it follows that the radius of the condensate halo only depends on the mass and scattering length of the dark matter particles. Note that the radius of the halo is independent of the central density.

The central density $\rho_c$ can be determined from the normalization condition, $\int d^3\boldsymbol{r}\rho=M$, thus obtaining
\be
\rho_c=\rho_0=\frac{Mk^3}{4\pi^2}=\frac{\pi M}{4R^3}=\frac{M}{4\pi^2}\left(\frac{Gm^3}{a\hbar ^2}\right)^{3/2}.
\ee

Once we have obtained from the hydrodynamic representation the condensate dark matter
density profile, all the global parameters of the halo
(radius, mass, central density), as well as the rotational velocities of particles in stable circular orbits can
be found in an exact form.  The knowledge of the physical quantities describing the Bose-Einstein Condensate dark matter halos gives us the important opportunity of the full observational test of this model.

\subsubsection{The mass of the dark matter particle}

The above astrophysical results allow us to obtain some simple estimates for one of the main characteristics of dark matter, the mass of its component particle. The use of Eq.~(\ref{rad}) leads to a first qualitative estimate of the
physical properties of the dark matter particle. The radius $R$ of the static condensate dark
matter halo is given, as a function of the mass $m$
and scattering length $a$ of the dark matter particle,  by $R=\pi \sqrt{\hbar ^{2}a/Gm^{3}}$. The
total mass of the condensate dark matter halo $M(R)$ is obtained, as a function of the radius and central density, as
\begin{equation}
M(R)=4\pi ^2\left(\frac{\hbar ^2a}{Gm^3}\right)^{3/2}\rho _c=\frac{%
4}{\pi}\rho _{c}R^3 .
\end{equation}
Moreover,  the mean value $\left<\rho \right>$ of the condensate density is given
by the expression $\left<\rho \right>=3\rho _{c}/\pi ^2$ \citep{BoHa07}.
Hence the mass of the dark matter particle in the Bose-Einstein Condensate is given by %
\citep{BoHa07}
\begin{eqnarray}  \label{mass}
m &=&\left( \frac{\pi ^{2}\hbar ^{2}a}{GR^{2}}\right) ^{1/3}
\notag \\
& \approx & 6.73\times 10^{-2} \left[ a\left( \mathrm{fm}\right) \right]
^{1/3}\left[ R\;\mathrm{(kpc)}\right] ^{-2/3}\;\mathrm{eV}.
\end{eqnarray}
If for the scattering length we adopt the value $a\approx 1 $ fm, and we take $R\approx 10$ kpc, the mass of the
condensate dark matter particle is of the order of $m\approx 14$ meV. If $%
a\approx 10^{6}$ fm,  a value of the scattering length corresponding to values observed in
terrestrial laboratory experiments, we obtain $m\approx 1.44$ eV.

 A number of important properties of dark matter can be obtained observationally from the study of the collisions between clusters of galaxies. Some well studied cases of such collisions are the
Bullet Cluster (1E 0657-56), and the Baby Bullet (MACSJ0025-12). From these
observations constraints on the physical properties of dark
matter can be obtained, including  the interaction cross-section between dark matter particles and baryonic matter, as well as the
dark matter-dark matter self-interaction cross section. If from the
observations the ratio $\sigma
_m=\sigma /m$ of the self-interaction cross section $\sigma =4\pi
a^2$ and of the dark matter particle mass $m$ is known, then with the use of Eq.~(\ref{mass}) the mass of the dark matter
particle in the Bose-Einstein condensate can be immediately obtained as
\begin{equation}
m\approx \left(\frac{\pi ^{3/2}\hbar ^2}{2G}\frac{\sqrt{\sigma _m}}{R^2}%
\right)^{2/5}.
\end{equation}

In \cite{Bul} an upper limit (68 \% confidence) for $\sigma
_m$ of the order of $\sigma _m<1.25\;\mathrm{cm^2/g}$ was found, through a comparison of the  results obtained  from weak lensing, strong lensing, optical and X-ray  observations, together with numerical simulations of the merging galaxy cluster 1E
0657-56 (the Bullet cluster).  By taking for $\sigma _m$ a numerical value of the order of $\sigma _m=1.25\;\mathrm{cm^2/g%
}$,  for the mass of the dark matter particle we obtain an upper limit of the
order of
\begin{eqnarray}\label{massdm}
& m < 3.1933\times10^{-37}\left(\frac{R}{10\;\mathrm{kpc}}%
\right)^{-4/5} \left(\frac{\sigma _m}{1.25\;\mathrm{cm^2/g}}\right)^{1/5}\;%
\mathrm{g}  \notag \\
& = 0.1791\times\left(\frac{R}{10\;\mathrm{kpc}}\right)^{-4/5} \left(\frac{%
\sigma _m}{1.25\;\mathrm{cm^2/g}}\right)^{1/5}\;\mathrm{meV}.
\end{eqnarray}

Note that the above mass limit is consistent with the limit obtained  in \cite{Bo} from a cosmological
analysis. For this value of the particle mass
the scattering length $a$ can be evaluated as
\begin{eqnarray}
a < \sqrt{\frac{\sigma _m\times m}{4\pi }} = 1.7827\times 10^{-6}\;%
\mathrm{fm}.
\end{eqnarray}

In \cite{Bul1} a stronger constraint for $\sigma _m$ was proposed, so that $%
\sigma _m\in(0.00335\;\mathrm{cm^2/g},0.0559\;\mathrm{cm^2/g})$. Hence we can constrain
the mass of the dark matter particle to a  range of values of the order of
\begin{eqnarray}
& m\approx \left(9.516\times 10^{-38}-1.670\times 10^{-37}\right)
\left(\frac{R}{10\;\mathrm{kpc}}\right)^{-4/5}\;\mathrm{g}  \notag \\
& = \left(0.053-0.093\right)\times \left(\frac{R}{10\;\mathrm{kpc}}%
\right)^{-4/5}\;\mathrm{meV}.
\end{eqnarray}
The scattering length can then be estimated to be in the range
\begin{eqnarray}
a \approx \left(5.038-27.255\right)\times 10^{-8}\;\mathrm{fm}.
\end{eqnarray}

Therefore the Bullet Cluster constraints and the galactic radii data predict
a condensate dark particle mass of the order of $m_{\chi }\approx 0.1$ meV.
In \cite{Bul2,Bul3} more precise results on the self-interacting dark matter cross section have been
obtained. From the
dark matter's lack of deceleration in the bullet cluster collision, and by using as tests of the non-gravitational forces acting on dark matter the collisions between galaxy clusters, for long-ranged forces
a self-interaction cross-section of the order of $\sigma_{m}/m <
1.25$ cm$^2$/g (68\% confidence limit) is obtained \cite{Bul2}.
A self-interaction cross-section $%
\sigma_{m}/m < 0.47$ cm$^2$/g (95\% CL) was inferred from the observation of 72 collisions \cite{Bul2}.

\subsection{The first order approximation for rotating dark matter halos}

In the following we will neglect the effects of the baryonic matter and of the random potential on the condensed Bose-Einstein dark matter halo. Then the general solution of Eq.~(\ref{dens}), giving the density profile of a pure rotating condensate can be represented as \cite{Zhang}
\begin{equation}\label{2cr}
  \rho(r,\theta)=\frac{\omega^2}{2\pi G}+\sum_{l=0}^{\infty}A_{2l}j_{2l}(kr)P_{2l}(\cos{\theta}),
\end{equation}
where $j_l(X)$ are the spherical Bessel functions, which are the solutions to the radial part of the Helmholtz equation. The solutions to the angular part of the Helmholtz equation are given by the Legendre polynomials $P_l(\cos{\theta})$. $A_{2l}$ are integration constants that must be determined generally from the continuity of the gravitational potential across the boundary of the dark matter halo. The total mass of the halo $M$ is defined as \cite{Zhang}
\be
M=2\pi \int_0^{2\pi}{\sin \theta d\theta}\int_0^{r_0}{\rho (r,\theta)r^2dr},
\ee
where $r_0$ is the radius of the galaxy. Note that by assuming that the dark matter halo rotates slowly, that is, $\omega ^2/G<<Mk^3$,  the deviation from  spherical symmetry can be considered small. Hence in the slow rotation approximation we may use perturbative method with a good approximation, and in order to perform the series expansions we  take the perturbation parameter as$ \sqrt{\omega ^2/GMk^3}$. Hence we can rewrite the density distribution (\ref{2cr}) as
\bea\label{3cr}
\rho (r,\theta)&=&\frac{\omega^2}{2\pi G}+\frac{Mk^3}{4\pi ^2}\Bigg[\frac{\sin kr}{kr}+\sum _{l=1}^{\infty}\left(\frac{\omega ^2}{GMk^3}\right)^{l/2}\times \nonumber\\
&&B_lj_{2l}(kr)P_{2l}(\cos{\theta})\Bigg],
\eea

The vacuum surface of the dark matter halo  $r_0=r_0(\theta)$ is defined by the condition $\rho (r,\theta)=0$. Since according to our basic assumption the deformation is small, $r_0$ may be expanded as
\be\label{4cr}
r_0-R=\sum _{l=1}^{\infty}{\left(\frac{\omega ^2}{GMk^3}\right)}^{l/2}C_{2l}(\theta).
\ee
Using Eqs.~(\ref{3cr}) and (\ref{4cr}), after  expanding $\rho \left(r_0 ,\theta \right)$ as a power series in $\omega ^2/GMk^3$, and equating the corresponding coefficients, we can obtain $C_{2l}(\theta )$. In the first order of approximation we obtain \cite{Zhang}
\begin{equation}\label{rho}
  \rho(r,\theta)=\frac{\omega^2}{2\pi G}+\left(\rho_c-\frac{\omega^2}{2\pi G}\right)j_0(kr)-\frac{5\omega^2\pi}{12G}j_2(kr)P_2(\cos{\theta}),
\end{equation}
or, equivalently,
\be\label{rhonew}
\frac{\rho(r,\theta)}{\rho _c}=\left(1-\Omega ^2\right)j_0(kr)+\Omega ^2\left[1-\frac{5\pi ^2}{6}j_2(kr)P_2(\cos{\theta})\right].
\ee

The vacuum boundary, giving the radius of the slowly rotating Bose-Einstein Condensate dark matter halo, can be computed as \cite{Zhang}
\bea\label{r_0}
\hspace{-0.6cm}  r_0(\theta)&=&\frac{\pi}k+\frac{\omega^2}{4G\rho_c k}\left[2-5P_2(\cos{\theta})\right]=\frac{\pi}{k}\times \nonumber\\
\hspace{-0.6cm} && \left\{1+\Omega ^2\left[1-\frac{5}{2}P_2(\cos \theta)\right]\right\},
\eea
while the equatorial radius of the dark matter halo is given by
\be
r_0\left(\frac{\pi}{2}\right)= \frac{\pi}{k}\left(1+\frac{9}{4}\Omega ^2\right).
\ee

In the non-rotating (static) case $\Omega =0$ we have $r_0\left(\pi/2\right)=R=\pi/k$. Hence the radius of the static Bose-Einstein Condensate dark matter halo is determined by the mass and scattering length of the dark matter particle only. For slowly rotating dark matter halos the mass distribution within a radius $r\leq \pi /k-3\omega^2/4G\rho_c k$ the mass profile is given by \cite{Zhang}
\bea\label{masseqn}
m(r)&=&\frac{4\pi \rho _c}{k^2}r\times \nonumber\\
&&\Bigg[\left(1-\Omega ^2\right)\frac{\sin kr}{kr}-\left(1-\Omega ^2\right)\cos kr +\Omega ^2\frac{(kr)^2}{3}\Bigg], \nonumber\\
\eea
where
\bea
\Omega ^2&=&\frac{\omega^2}{2\pi G\rho _c}=0.02386\times \nonumber\\
&&\times \left(\frac{\omega }{10^{-16}\;{\rm s}^{-1}}\right)^2\times \left(\frac{\rho _c}{10^{-24}\;{\rm g/cm^3}}\right)^{-1}.
\eea

Note that as compared to the non-rotating case, a second term, due to the presence of the rigid body type rotation, comes out in the mass profile. The tangential velocity $v_{tg}^2$ of massive test particles rotating in the Bose-Einstein Condensate galactic dark matter halo is calculated immediately as
\be
v_{tg}^2=\frac{Gm(r)}{r}.
\ee

In the first order of approximation $v_{tg}^2$ is obtained as
\bea
&&v_{tg}^2(r)=\frac{4G\rho_cR^2}{\pi}\times \nonumber\\
&&\Bigg[\left(1-\Omega ^2\right)\frac{\sin \left(\pi r/R\right)}{\pi r/R}-\left(1-\Omega ^2\right)\cos \frac{\pi r}{R} +\frac{\Omega ^2}{3}\left(\frac{\pi r}{R}\right)^2\Bigg], \nonumber\\
\eea
or, equivalently \footnote{This equation must replace Eq.~(103) of reference \cite{Zhang}, where the term proportional to $\left(\rho _c/10^{-24}\;{\rm g/cm^3}\right)^{-1}$ in the definition of the tangential velocity is missing.},
\bea
\hspace{-1.5cm}&&v_{tg}^2\left( \;{\rm km^2/s^2}\right)= 80.861\times \left(\frac{\rho _c}{10^{-24}\;{\rm g/cm^3}}\right)\times \left(\frac{R}{{\rm kpc}}\right)^2\times \nonumber\\
\hspace{-1.5cm}&&\Bigg[\left(1-\Omega ^2\right)\frac{\sin \left(\pi r/R\right)}{\pi r/R}-\left(1-\Omega ^2\right)\cos \frac{\pi r}{R} +\frac{\Omega ^2}{3}\left(\frac{\pi r}{R}\right)^2\Bigg], \nonumber
\label{eq_vh}
\\
\eea
where we have introduced the radius $R=\pi/k$ of the static galactic halo.

\section{Comparing the galactic rotation curves in the Bose-Einstein Condensate dark matter model with the SPARC data}\label{sect2}

In the present Section we will perform a detailed comparison of the theoretical predictions of the Bose-Einstein Condensate dark energy model with the observations of the galactic rotation curves. In our study we will use the data provided by the new Spitzer Photometry and Accurate Rotation Curves (SPARC) database \cite{Lelli1,Lelli2}. After performing a number of cuts in the data, we will fit the theoretical expression of the tangential velocity given by Eq.~(\ref{eq_vh}) with the SPARC data. The basic parameters of the dark matter halos (central density, rotational velocity, radius etc.) are obtained from the fitting, and the density profile of the dark matter is also presented. The estimations of the mass of the condensate dark matter halo are also presented.

\subsection{The SPARC sample}

 The SPARC sample contains the data of 175 galaxies, with all rotationally supported morphological types present. The database covers an extensive
  interval in galactic rotation velocities, Hubble types, luminosities, and surface brightness, respectively. It also includes near-infrared (3.6$\mu$m) observations that account for the distribution of stellar masses, as well as 21 cm observations that trace the atomic gas. The rotation curves are derived from the 21 cm velocity fields. High spatial resolution observations of ionized interstellar supplements these data.  To date SPARC is the largest galactic database that for every galaxy contains rotation curves as well as spatially resolved data on the distribution of both gas and stars.

  The SPARC dataset includes galaxies having rotation
velocities in the range $20 < Vf < 300$ km/ s, luminosities  $10^7 <L_{[3.6]} < 5 \times  10^{11} L_{\odot}$, gas masses in the range $10^7 < M_{gas} < 5 \times 10^{10} M_{\odot}$, gas fractions $0.01 < F_{gas} < 0.97$, effective surface brightnesses
$5 < \Sigma _{*} < 3 \times 10^3 L_{\odot}$ /pc$^{2}$,  and half-light radii $0.3 < R_{1/2} < 5$ kpc, respectively \cite{Lelli1,Lelli2,Lelli3}. Note that the data contain information about  some of the largest known individual galaxies and about many  small ones. Low surface brightness and low mass   galaxies are very well represented in SPARC. This is in opposition to flux selected samples, which are generally limited to the interval $M_{*} > 10^9 M_{\odot}$ and $V_f > 100$ km/s \cite{Lelli3}.

In Fig.~\ref{fig_gal_type} we plotted the number of galaxies of each type present in the SPARC sample.
\begin{figure}[h]
\begin{center}
\includegraphics[]{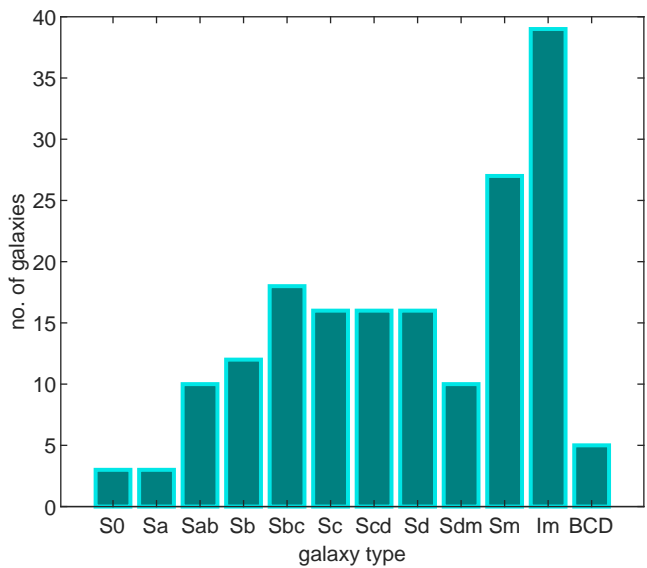}
\caption{\label{fig_gal_type}\small{The number of galaxies of each type of the SPARC sample.
}}
\end{center}
\end{figure}

In Fig.~\ref{fig_data_len} we present a histogram of the number of data points (lengths) for the galactic velocities.
\begin{figure}[h]
\begin{center}
\includegraphics[scale=0.65]{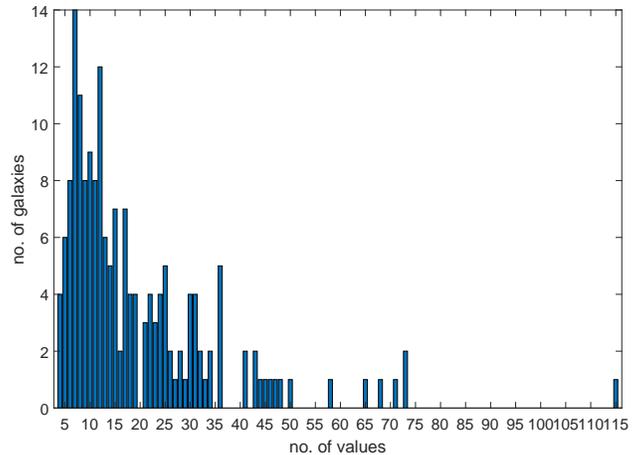}
\caption{\label{fig_data_len}\small{Histogram of the number of data points of the rotation velocity curves in the SPARC sample.
}}
\end{center}
\end{figure}

We have fitted our theoretical  model using only galaxies having  lengths greater than or equal to 7. Therefore we have restricted from the beginning our analysis to a number of $157$ galaxies of the SPARC sample.  In the SPARC sample a number of $143$ galaxies do not contain any data on the contribution of the stellar bulge to the baryonic rotation curve. From these galaxies 125 have more than 6 observation points. There are a total of 32 galaxies containing data on the stellar bulge, all these galaxies having a length of the data greater than 6.

\subsection{Material and methods}\label{secMM}

In order to compare the velocity of the test particles in the Bose-Einstein Condensate dark matter halo, as given by Eq. (\ref{eq_vh}), we have fitted it with the total observational velocity of the particles that can be obtained from the SPARC database,
and which can be obtained from the relation
\begin{equation}
v_t=\sqrt{v_{gas}|v_{gas}|\!+\!\Upsilon_d\!\times\! v_{disk}|v_{disk}|\!+\!\Upsilon_b\!\times\! v_{bulge}|v_{bulge}|\!+\!v_{tg}^{2}},
\label{v_tot}
\end{equation}
where $v_t$ is the total velocity, including the contributions of both baryonic and dark matter, and $v_{gas}$, $v_{disk}$, $v_{bulge}$ and $v_{tg}$ denote the contributions from the velocity of the gas, of the disk, of the bulge, and of the dark matter halo, respectively. $\!\Upsilon_d\!$ and $\!\Upsilon_b\!$ denote the  stellar mass-to-light ratios for the disk and the stellar bulge, respectively.

To perform the fitting, and by taking into account the observational errors,  we have looked for the minimum of the objective function
\begin{equation}
\chi^2(R,\rho _{c},\omega,\Upsilon_d,\Upsilon_b)=\frac{1}{n-k}\sum_{i=1}^{n}\frac{%
(v_{t,i}-v_{obs,i})^{2}}{\sigma _{i}^{2}},
\label{chi2}
\end{equation}
where $n$ is the length of data, and $k$ is the number of parameters that must be estimated, that is, $k=4$ for bulgeless galaxies and $k=5$ for galaxies with bulge, respectively.

We have limited the values of $\Upsilon_d$ and $\Upsilon_b$ to the interval [0.1,5].
We have also imposed the physical constraint that the dark matter density must be a monotonically decreasing function of the radial coordinate, and thus $\rho$ must decrease for all $r<r_{max}$. At the boundary of the dark matter halo the density must vanish, and this condition fixes the radius $R_h$ of the halo. For the rotation velocity of the galactic halo we have imposed the constraint  $\Omega^2<0.109$ , while for the lower bound of the central density of the dark matter $\rho_c$ we have adopted the value $10^{-27} {\rm g/cm^{3}}$.  As for the upper bound for the static radius $R$ of the dark matter halo, we have taken the value $r_{max}/2$.

\subsubsection{The fitting algorithm}

In order to find the parameters that minimizes the objective function under the above mentioned constraints we have used genetic algorithms.
The genetic algorithms are inspired by the process of natural selection (`survival of the fittest'), and they  belong to the larger class of evolutionary algorithms. They are commonly used to generate high-quality solutions to optimization and search problems by relying on bio-inspired operators such as mutation, crossover and selection.

In order to use genetic algorithms one must define a fitness function (in our case  this is the objective function defined by Eq.~(\ref{chi2}).
The first step of the genetic algorithms is initialization, which means that we consider an initial population of random candidates for solutions (in a specified range), and compute the fitness value for each candidate. Then iteratively we generate new populations  by selection, crossover and mutation, until a stopping criterion is satisfied. Elitist selection methods rate the fitness of each solution and preferentially select the best solutions.

Crossover (or recombination)- is a genetic operator used to combine the genetic information of two parents to generate new offsprings. In our case it recombines sets of parameters in order to obtain new better sets. We have mostly used an heuristic crossover that returns a child that lies on the line containing the two parents, a small distance away from the parent with the better fitness value in the direction away from the parent with the worse fitness value.
Mutation alters one or more values of parameters. We have used a mutation of Gaussian type
that makes small random changes in the parameters in order to obtain the parameters of the next population.

We have chosen for the total population size values between [4000,5000], for crossover fraction (the ratio from the population size that will be recombined for the next population) values in [0.6,0.9]. The number of the sets of parameters with the lowest values of the fitness function that will be kept unchanged from one population to another was chosen between $1\%$ and $5\%$ from the total population size. The initial population range was adapted for each galaxy.
 The  maximum number of  generations was set to $100$. In Fig.~\ref{fig_pop_evol_NGC0055} one can see how the parameters $R$ and $\rho_c$ evolves toward the minimum. At latter iterations most of the population concentrates around the minimum.

\begin{figure}[tbp]
\begin{center}
\includegraphics[scale=0.65]{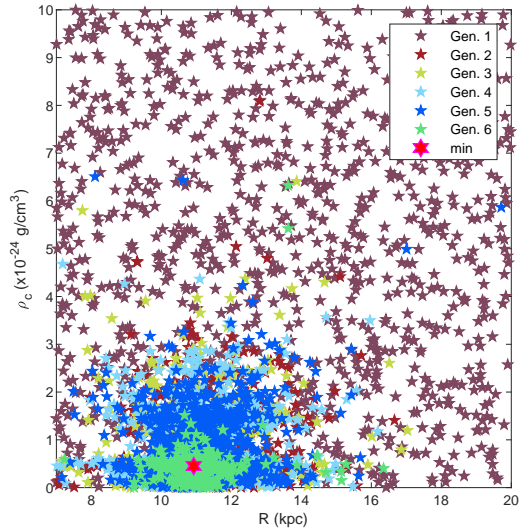}
\caption{\label{fig_pop_evol_NGC0055}\small{The evolution of the two first parameters $R$ and $\rho_c$ of the population at different iterations toward the minimum.
}}
\end{center}
\end{figure}

\subsection{Fitting results}

In order to test the Bose-Einstein Condensate dark matter model we have performed a detailed analysis of the theoretical model with the SPARC sample. The basic observational quantities for testing any dark matter model are the galactic rotation curves. The BEC model also gives some definite predictions on the dark matter distribution, which can be obtained from the basic theory.

\subsubsection{The galactic rotation curves}

In order to test the BEC model with the SPARC data we have to consider two classes of galaxy data: those who does not contain information on the effects of the stellar bulge on the baryonic contribution to the total velocity, and those which do show these contributions. A selected sample of rotation curves for bulgeless galaxies, fitted with the theoretical curve of the BEC model, is presented in Fig.~\ref{fig_viteze_gal1}. The rotational curves corresponding to the optimal parameters obtained for the sample of galaxies without bulge data (listed in Table~\ref{table1}) are plotted with thick continuous magenta line.


\begin{figure*}[tbp]
\begin{center}
\includegraphics[width=\textwidth]{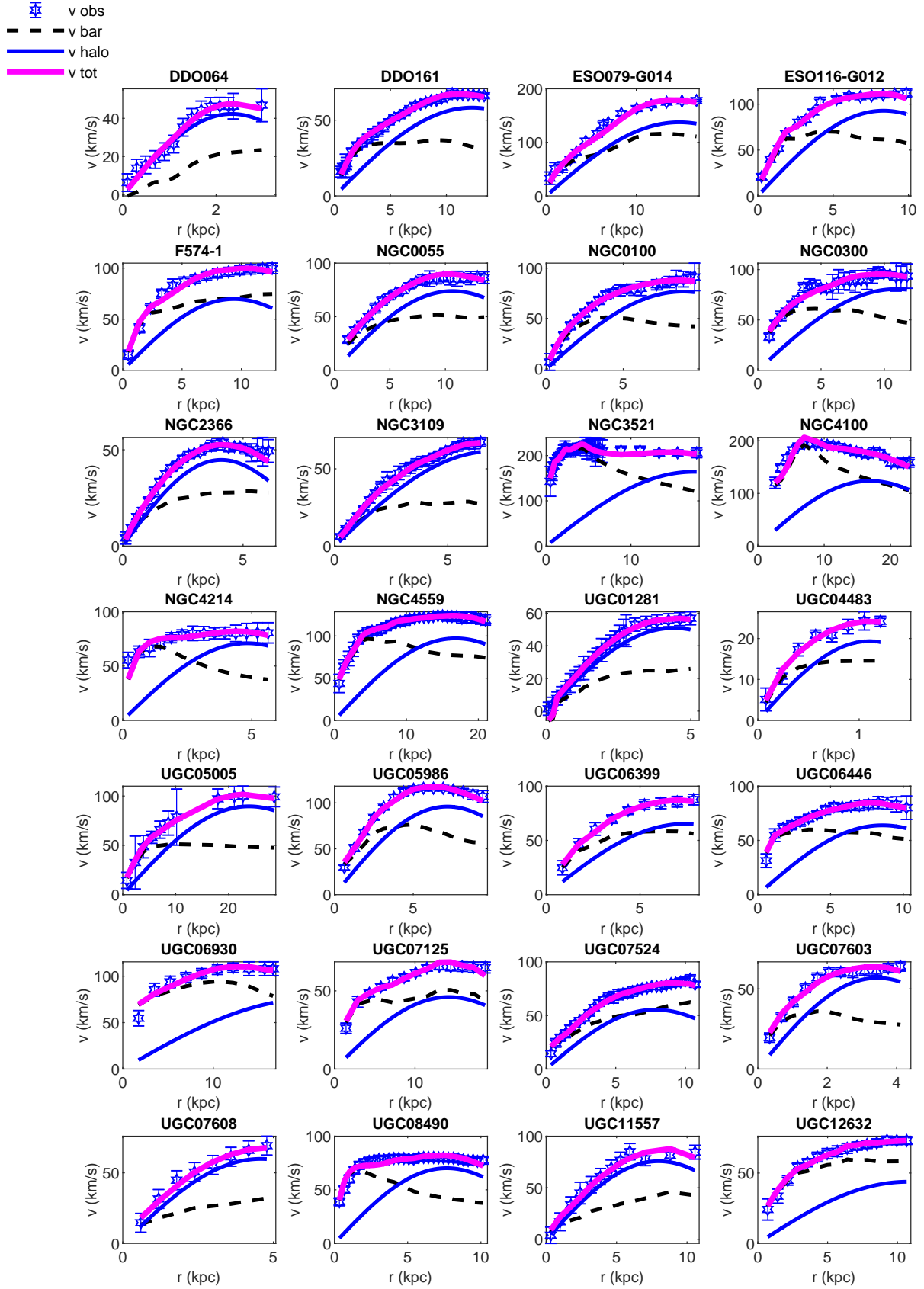}
\vspace{-2cm}
\caption{\label{fig_viteze_gal1} The rotation velocity for a selected sample of bulgeless galaxies. The observed total velocity is plotted with error bars, the estimated baryonic contributions (disk and gas contributions) with black dashed lines, while the estimated total velocity is plotted with continuous magenta line.}
\end{center}
\end{figure*}

Fig.~\ref{fig_viteze_gal_vbulge}  represents the observed rotation velocities and the best-fit curves for a selected sample of galaxies with stellar bulge data.
For these galaxies we had five parameters to optimize, and the optimal values of the parameters are listed in Table~\ref{table2}.

\begin{figure*}[tbp]
\begin{center}
\includegraphics[width=\textwidth]{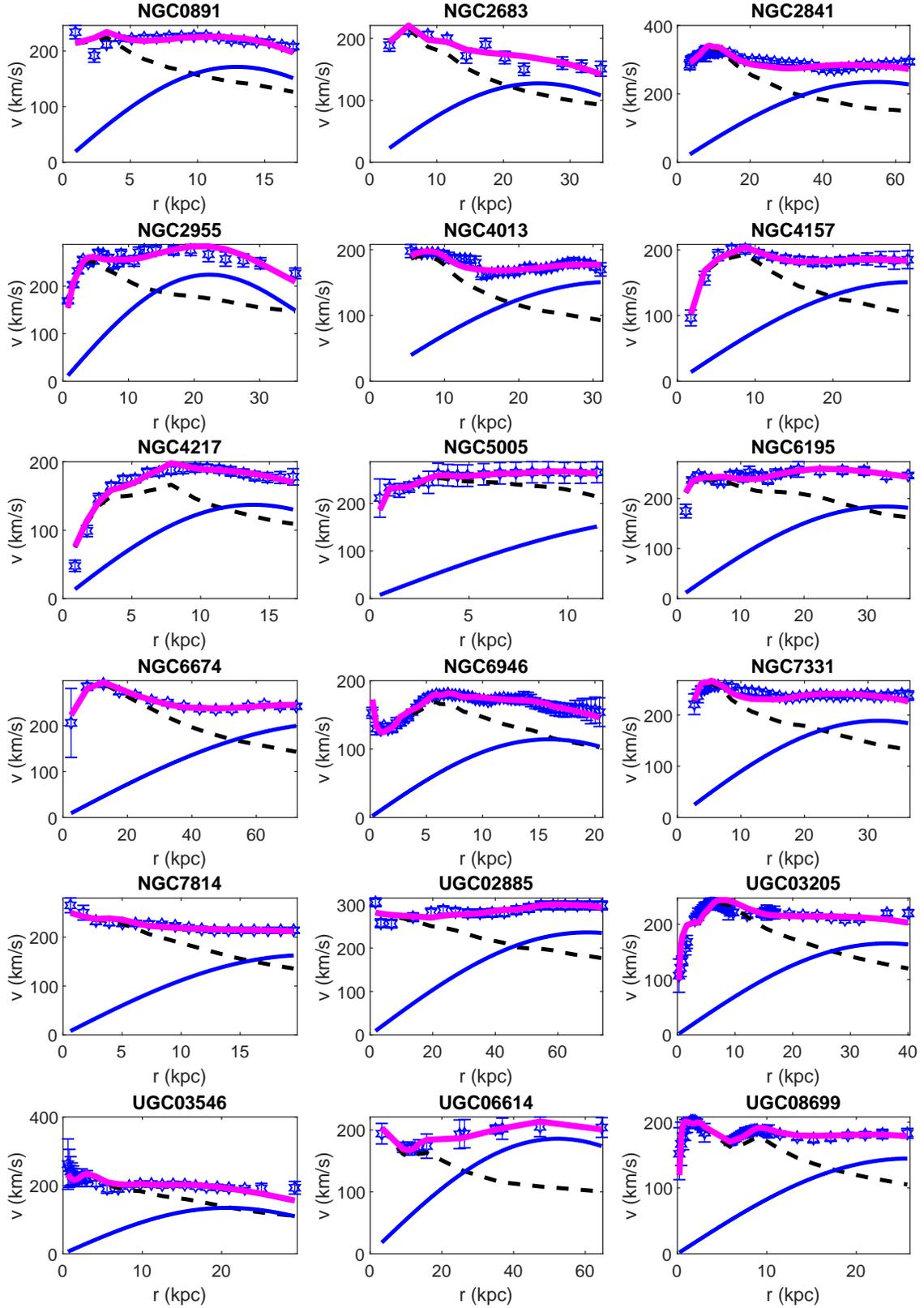}
\vspace{-1cm}
\caption{\label{fig_viteze_gal_vbulge}\small{The rotational velocities for a selected sample of  galaxies with stellar bulge data.
}}
\end{center}
\end{figure*}

The third group of galaxies we have considered are those whose angular velocity can be roughly considered as zero. Fig.~\ref{fig_viteze_omega0} shows the best-fit rotation curves for a number galaxies for which the estimated angular velocity $\omega$ is zero, corresponding to the non-rotating case. In this situation the static BEC model gives the description of the static and dynamic properties of the dark matter halo, and $R_h=R$. The optimal parameters for these galaxies are displayed in Table~\ref{table3}.

\begin{figure*}[tbp]
\begin{center}
\includegraphics[width=\textwidth]{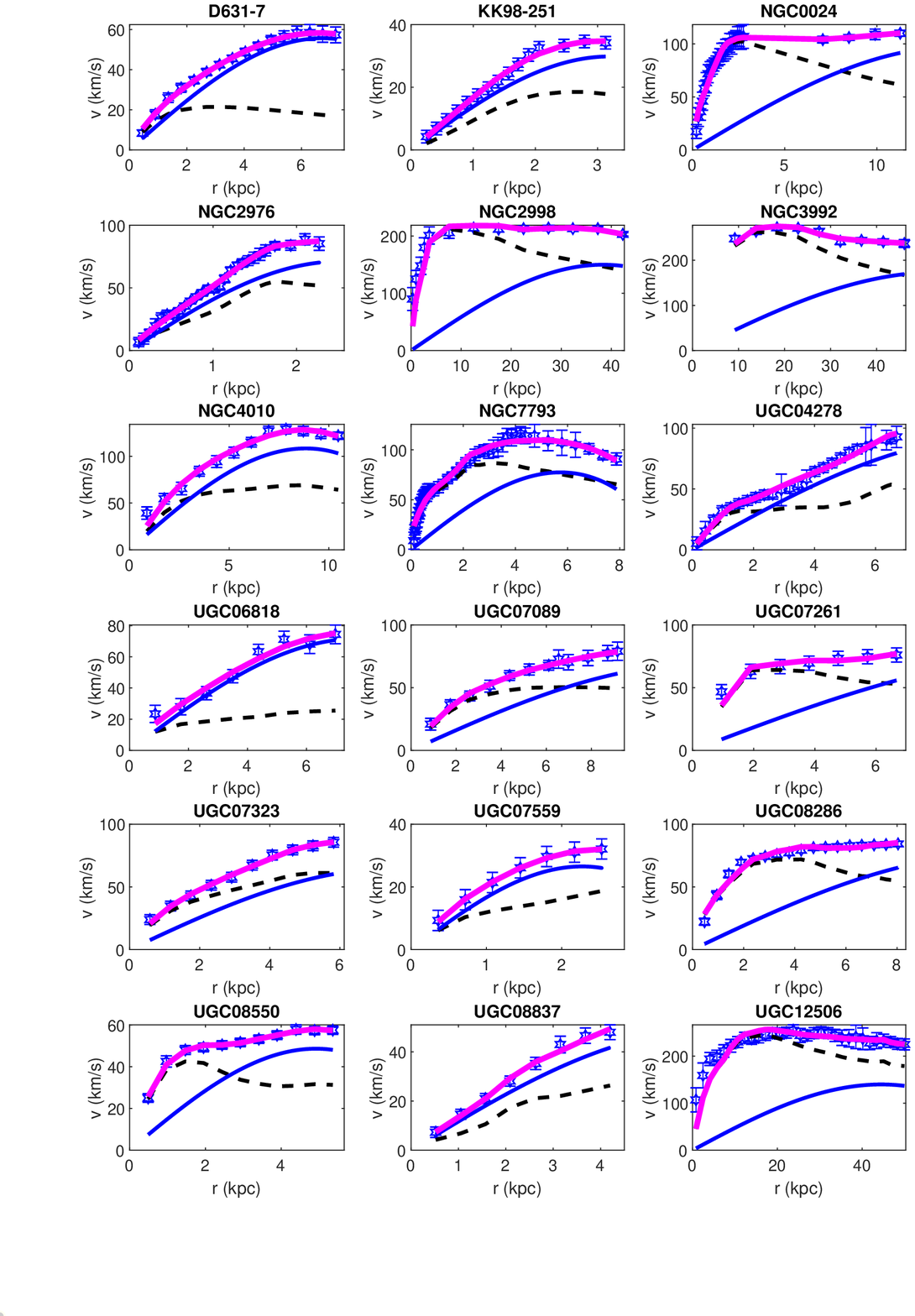}
\vspace{-1cm}
\caption{\label{fig_viteze_omega0}\small{The rotation velocity for galaxies for which the estimated dark matter halo angular velocity is zero}}
\end{center}
\end{figure*}

\subsubsection{The density profiles}

Once the physical parameters of the BEC dark matter halos are known from the fitting of the rotation curves, we can obtain all the physical parameters of the dark matter halos. One such parameter is the dark matter density distribution. In Fig.~\ref{fig_density} we have displayed the density profiles of the dark matter halos for the twenty eight  (bulgeless) galaxies, whose rotation velocity curves are represented in Fig.~\ref{fig_viteze_gal1}. With vertical dashed lines we pointed the values of the estimated static and rotating  radius of the dark matter halo $R$, and $R_h$, respectively. For all considered cases the condition $R_h>R$ holds, indicating that rotation leads to an increase of the radius of the galactic halo.

\begin{figure*}[tbp]
\begin{center}
\includegraphics[scale=0.7]{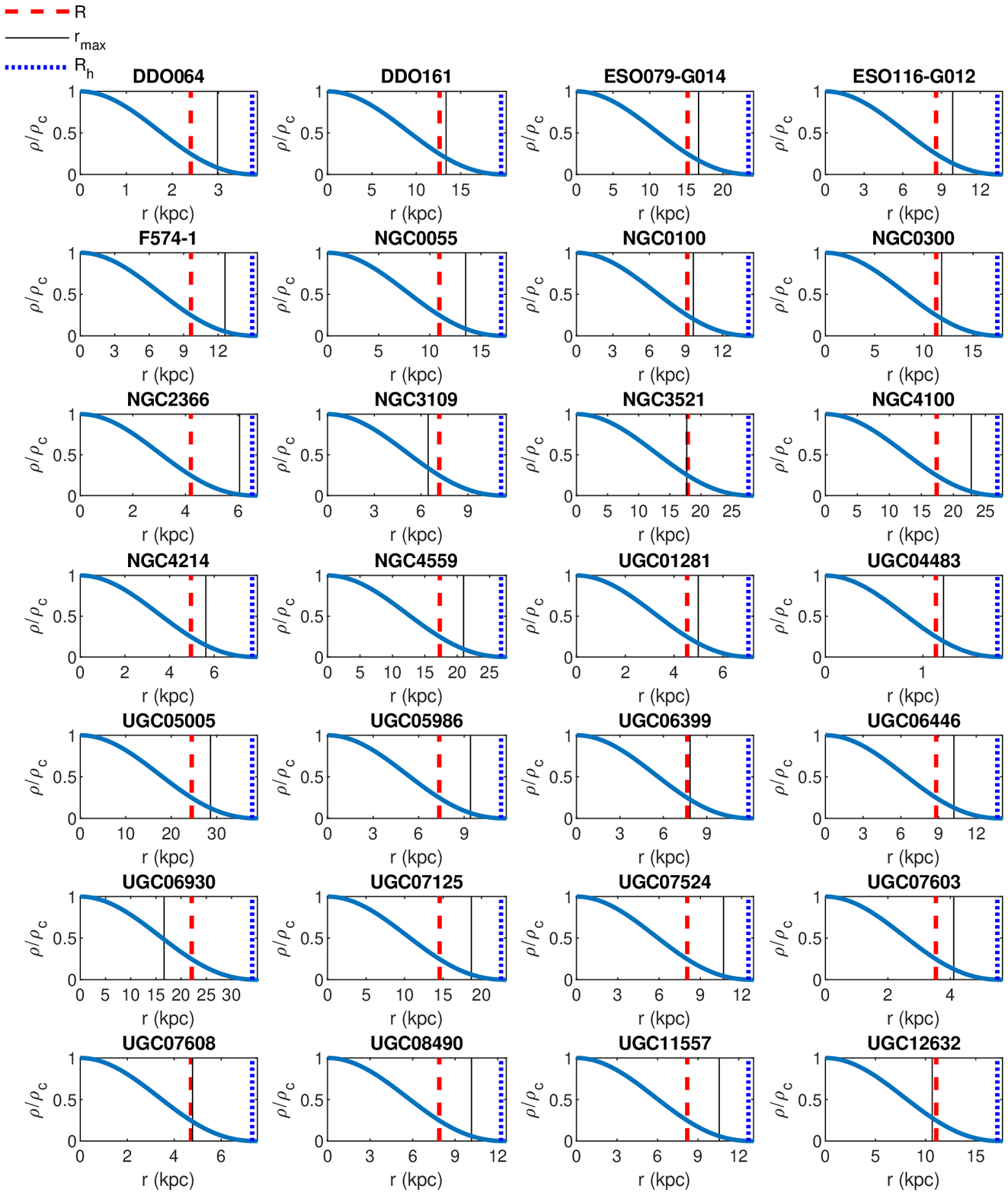}
\caption{\label{fig_density}\small{The predicted density profiles of the Bose-Einstein Condensate dark matter halos for the galaxies whose velocity curves are plotted in Fig.~\ref{fig_viteze_gal1}. With red dashed line and blue dotted line are represented the values of the estimated $R$, respectively $R_h$, while with black continuous line is represented the maximum value of $r$ (for which we have observed data).}}
\end{center}
\end{figure*}

\begin{table}
\small  
\centering
\resizebox{\columnwidth}{!}{

\begin{tabular}{|l|r|r|c|c|c|c|}

\hline

Galaxy & $R\quad$ & $R_h\quad$ &$   \rho_c  $ & $\omega $ & $\Upsilon_d $ &   $\chi^2$ \\
\  & ({\rm kpc}) &  ({\rm kpc}) & $(10^{-24}\; {\rm g/cm^{3}})$ & $(10^{-16}\; {\rm s}^{-1})$ &  $ (M_\odot/L_\odot)$   & \  \\

\hline
DDO064 & 2.401 & 3.736 & 3.067 & 3.756 & 0.190 & 0.336 \\
\hline
DDO161 & 12.622 & 19.638 & 0.210 & 0.983 & 0.868 & 0.319 \\
\hline
ESO079-G014 & 15.171 & 23.605 & 0.810 & 1.930 & 0.692 & 2.544 \\
\hline
ESO116-G012 & 8.581 & 13.351 & 1.154 & 2.304 & 1.012 & 1.027 \\
\hline
F574-1 & 9.637 & 14.993 & 0.516 & 1.541 & 1.850 & 0.862 \\
\hline
NGC0055 & 10.925 & 16.998 & 0.451 & 1.441 & 0.577 & 0.301 \\
\hline
NGC0100 & 9.114 & 14.180 & 0.698 & 1.792 & 0.814 & 0.218 \\
\hline
NGC0300 & 11.255 & 17.512 & 0.506 & 1.525 & 1.415 & 0.612 \\
\hline
NGC2366 & 4.206 & 6.545 & 1.116 & 2.265 & 0.471 & 0.299 \\
\hline
NGC3109 & 7.167 & 11.151 & 0.723 & 1.823 & 2.320 & 0.248 \\
\hline
NGC3521 & 17.904 & 27.856 & 0.830 & 1.953 & 0.586 & 0.165 \\
\hline
NGC4100 & 17.366 & 27.019 & 0.500 & 1.517 & 0.851 & 1.521 \\
\hline
NGC4214 & 4.956 & 7.712 & 2.025 & 3.052 & 1.334 & 0.818 \\
\hline
NGC4559 & 17.313 & 26.937 & 0.311 & 1.196 & 0.669 & 0.069 \\
\hline
UGC01281 & 4.533 & 7.052 & 1.246 & 2.393 & 0.767 & 0.320 \\
\hline
UGC04483 & 1.133 & 1.762 & 2.876 & 3.637 & 1.139 & 0.551 \\
\hline
UGC05005 & 24.495 & 38.112 & 0.132 & 0.778 & 1.109 & 0.045 \\
\hline
UGC05986 & 7.360 & 11.452 & 1.678 & 2.778 & 1.032 & 0.841 \\
\hline
UGC06399 & 7.647 & 11.898 & 0.719 & 1.818 & 1.506 & 0.167 \\
\hline
UGC06446 & 8.803 & 13.697 & 0.519 & 1.545 & 3.241 & 0.208 \\
\hline
UGC06930 & 22.095 & 34.378 & 0.117 & 0.734 & 1.718 & 0.786 \\
\hline
UGC07125 & 14.562 & 22.658 & 0.099 & 0.673 & 1.059 & 0.974 \\
\hline
UGC07524 & 8.047 & 12.521 & 0.467 & 1.466 & 1.798 & 0.458 \\
\hline
UGC07603 & 3.540 & 5.508 & 2.547 & 3.423 & 1.079 & 0.923 \\
\hline
UGC07608 & 4.710 & 7.329 & 1.595 & 2.709 & 1.675 & 0.232 \\
\hline
UGC08490 & 7.880 & 12.260 & 0.782 & 1.897 & 2.081 & 0.787 \\
\hline
UGC11557 & 8.193 & 12.748 & 0.844 & 1.970 & 0.112 & 0.510 \\
\hline
UGC12632 & 11.055 & 17.200 & 0.155 & 0.844 & 3.404 & 0.242 \\
\hline

\end{tabular}
}
\caption{The optimal parameters for the bulgeless galaxies whose velocity curves are plotted in Fig.~\ref{fig_viteze_gal1}.}
\label{table1}
\end{table}

\begin{table}
\centering
\resizebox{\columnwidth}{!}{

\begin{tabular}{|l|r|c|c|c|c|c|c|}
\hline
Galaxy & $R\quad $ & $R_h\quad $ & $\rho_c  $ & $\omega $ & $\Upsilon_d $ & $\Upsilon_b$  &  $\chi^2$ \\
\  & ({\rm kpc}) & ({\rm kpc}) & $(10^{-24}\; {\rm g/cm^{3}})$ & $(10^{-16}\; {\rm s}^{-1})$ & $ (M_\odot/L_\odot)$ & $ (M_\odot/L_\odot)$   & \  \\
\hline
NGC0891 & 13.276 & 20.568 & 1.648 & 2.753 & 0.331 & 0.625 & 3.589 \\
\hline
NGC2683 & 25.929 & 39.984 & 0.238 & 1.047 & 0.873 & 0.100 & 2.520 \\
\hline
NGC2841 & 56.263 & 86.884 & 0.172 & 0.889 & 1.573 & 0.747 & 3.955 \\
\hline
NGC2955 & 22.928 & 35.379 & 0.942 & 2.082 & 0.258 & 0.864 & 2.784 \\
\hline
NGC4013 & 33.484 & 51.899 & 0.201 & 0.960 & 0.595 & 1.188 & 0.822 \\
\hline
NGC4157 & 30.892 & 47.936 & 0.235 & 1.041 & 0.574 & 0.100 & 0.251 \\
\hline
NGC4217 & 14.202 & 21.980 & 0.920 & 2.057 & 1.478 & 0.206 & 1.754 \\
\hline
NGC5005 & 18.238 & 27.904 & 0.938 & 2.072 & 0.606 & 0.499 & 0.061 \\
\hline
NGC6195 & 33.771 & 52.361 & 0.293 & 1.160 & 0.360 & 0.779 & 1.668 \\
\hline
NGC6674 & 97.994 & 98.008 & 0.051 & 0.012 & 1.442 & 0.948 & 1.378 \\
\hline
NGC6946 & 16.323 & 25.244 & 0.483 & 1.491 & 0.625 & 0.540 & 1.578 \\
\hline
NGC7331 & 32.577 & 50.525 & 0.331 & 1.234 & 0.493 & 0.100 & 2.609 \\
\hline
NGC7814 & 21.578 & 33.307 & 0.567 & 1.615 & 2.079 & 0.496 & 0.828 \\
\hline
UGC02885 & 71.298 & 109.952 & 0.108 & 0.705 & 0.895 & 1.014 & 1.326 \\
\hline
UGC03205 & 37.434 & 57.988 & 0.192 & 0.941 & 0.972 & 1.141 & 4.201 \\
\hline
UGC03546 & 21.275 & 32.842 & 0.393 & 1.345 & 0.898 & 0.319 & 1.568 \\
\hline
UGC06614 & 53.704 & 82.958 & 0.118 & 0.737 & 0.874 & 0.644 & 0.193 \\
\hline
UGC08699 & 26.621 & 41.045 & 0.291 & 1.158 & 1.321 & 0.535 & 1.094 \\
\hline
\end{tabular}
}
\caption{The optimal parameters for the galaxies with stellar bulge whose velocity curves are plotted in Fig.~\ref{fig_viteze_gal_vbulge}.}
\label{table2}
\end{table}

\begin{table}
\small  
\centering
\resizebox{\columnwidth}{!}{
\begin{tabular}{|l|r|c|c|c|c|}

\hline

Galaxy & $R\quad$ & $   \rho_c  $ & $\omega $ & $\Upsilon_d $ &   $\chi^2$ \\
\  & ({\rm kpc}) &  $(10^{-24}\; {\rm g/cm^{3}})$ & $(10^{-16}\; {\rm s}^{-1})$ &  $ (M_\odot/L_\odot)$   & \  \\
\hline
D631-7 & 7.701 & 0.608 & 0.000 & 0.100 & 0.246 \\
\hline
KK98-251 & 3.657 & 0.774 & 0.000 & 0.100 & 0.201 \\
\hline
NGC0024 & 16.429 & 0.409 & 0.000 & 2.087 & 0.339 \\
\hline
NGC2976 & 2.886 & 7.063 & 0.000 & 0.332 & 0.293 \\
\hline
NGC2998 & 43.916 & 0.136 & 0.000 & 0.912 & 2.446 \\
\hline
NGC3992 & 61.324 & 0.094 & 0.000 & 1.133 & 1.227 \\
\hline
NGC4010 & 10.125 & 1.334 & 0.000 & 0.368 & 0.904 \\
\hline
NGC7793 & 6.554 & 1.623 & 0.000 & 0.744 & 0.541 \\
\hline
UGC04278 & 11.658 & 0.747 & 0.000 & 1.134 & 0.384 \\
\hline
UGC06818 & 8.865 & 0.760 & 0.000 & 0.177 & 1.109 \\
\hline
UGC07089 & 14.959 & 0.249 & 0.000 & 0.838 & 0.141 \\
\hline
UGC07261 & 14.814 & 0.324 & 0.000 & 1.559 & 1.506 \\
\hline
UGC07323 & 8.710 & 0.647 & 0.000 & 0.785 & 0.245 \\
\hline
UGC07559 & 2.601 & 1.210 & 0.000 & 0.329 & 0.020 \\
\hline
UGC08286 & 13.891 & 0.352 & 0.000 & 3.071 & 1.779 \\
\hline
UGC08550 & 5.627 & 0.868 & 0.000 & 2.037 & 0.139 \\
\hline
UGC08837 & 7.316 & 0.522 & 0.000 & 0.250 & 0.380 \\
\hline
UGC12506 & 50.963 & 0.088 & 0.000 & 1.860 & 1.092 \\
\hline
UGCA442 & 6.712 & 0.660 & 0.000 & 3.537 & 0.617 \\
\hline
\end{tabular}
}
\caption{The optimal parameters for the nonrotating galaxies with $\omega =0$  whose velocity curves are plotted in Fig.~\ref{fig_viteze_omega0}.}
\label{table3}
\end{table}

\subsubsection{Statistical results}

There are $157$ galaxies having data length greater than or equal to 7 in the SPARC database, and which we have considered in our study.  From the fitting of the SPARC rotation curves we have obtained a value of $\chi^2$ less than $5$ and reasonable values for the estimated parameters (under the constraints imposed on the parameters mentioned in Subsection ~\ref{secMM}) for $117$ galaxies. In fact,  $\chi^2$ was less than $0.5$ for $41$ galaxies, less than $1$ for $61$ galaxies, less than $2$ for $88$ galaxies, and less than $3$ for $105$ galaxies.

In Fig.~\ref{fig_chi2_values} we have plotted the histogram of all the values of $\chi^{2}$ for the galaxies with reasonable parameters, while in the zoomed in picture we have plotted the histogram for the values smaller than 3.

\begin{figure}[h]
\begin{center}
\includegraphics[width=\columnwidth]{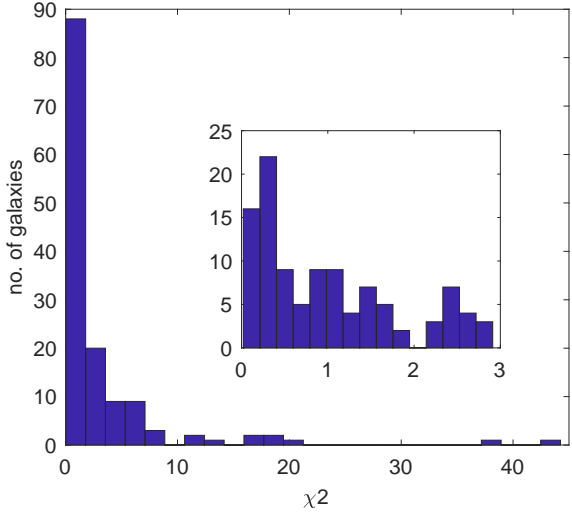}
\caption{\label{fig_chi2_values}\small{The histogram of the values of the objective function $\chi^{2}$; in the zoomed in picture is the histogram of the values smaller than 3}
}
\end{center}
\end{figure}

\subsection{Physical parameters of BEC dark matter halos}

 \paragraph {The density profile.} With the physical and astrophysical parameters of the BEC dark matter halos obtained from the fitting of the rotation curves we can now proceed to discuss the global properties of the galaxies, and try to infer the properties of the dark matter particle. The first important result of our analysis, and which follows directly from the theoretical model, is that the density distribution of the BEC dark matter is nonsingular at the galactic center, thus indicating the existence of a cored central profile. It has already been suggested \citep{Har1} that the Bose-Einstein Condensation of dark matter can solve the long standing core-cusp problem of dark matter distribution, and hence eliminate the unphysical properties that appear in the Navarro-Frenk-White density profile \cite{cusp}. Our results, based on the detailed analysis of the SPARC sample, also clearly predict the presence of a core at the galactic center, a finding which is supported by the astronomical observations. Near the galactic center, by ignoring the effects of the rotation, the density distribution can be approximated as
 \be
 \rho (r)\approx \rho _c\left(-\frac{1}{6} r^2 k^2 +\frac{1}{120} k^4 r^4
  \right)+O\left(r^5\right).
 \ee

 Another important and specific property of the BEC dark matter halos is their finite extension, with the dark matter density becoming effectively zero at a given radius $R_h$. This property generally does not appear in most of the other dark matter profiles considered in the literature. The radius of the BEC distribution depends on two parameters, the radius of the static configuration, and on the global rotation of the halo.

 \paragraph{The central density distribution.} An important fitting parameter, the central density $\rho _c$ of the condensate dark halo, has also been obtained from our analysis. The numerical values of the central density of the dark matter varied generally between around $\rho _c\approx 0.1\times 10^{-24}\;{\rm g/cm^3}$ and $\rho _c\approx 3\times 10^{-24}\;{\rm g/cm^3}$, but in a few cases larger central densities have also been obtained. The statistical distribution of the estimated values of the central density $\rho_c$ obtained for the entire SPARC sample is represented in Fig.~\ref{fig_hist_rho}. Obtaining the central density of the dark matter at the galactic center may prove to be an extremely difficult observational task,  but, if such a possibility may exist, the central density distribution of the dark matter may provide, at least in principle, a specific signature that may help to discriminate BEC dark matter models from other theoretical dark matter models.

\begin{figure}[h]
\begin{center}
\includegraphics[scale=0.6]{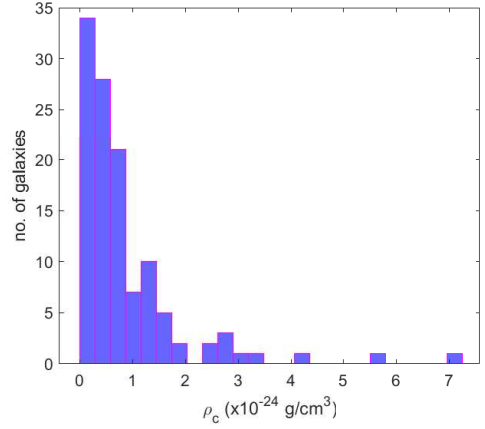}
\caption{\label{fig_hist_rho}\small{The distribution of the estimated values of the central density $\rho_c$ of the condensate dark matter halos.
}}
\end{center}
\end{figure}

Fig.~\ref{fig_R_ro} shows the values of the estimated  central density $\rho_c$ versus the values of the estimated static radius of the dark matter $R$. One can observe that they are slightly anti-correlated, the Pearson correlation coefficient \cite{Craciun} having the value $-0.429$ for the statistics 1, and $-0.463$ for the reduced statistics. In fact, from a physical point of view one can conclude that a larger radius of the dark matter with a smaller central density has similar influence on the halo velocity as a smaller radius and a larger central density.

\begin{figure}[h]
\begin{center}
\includegraphics[width=\columnwidth]{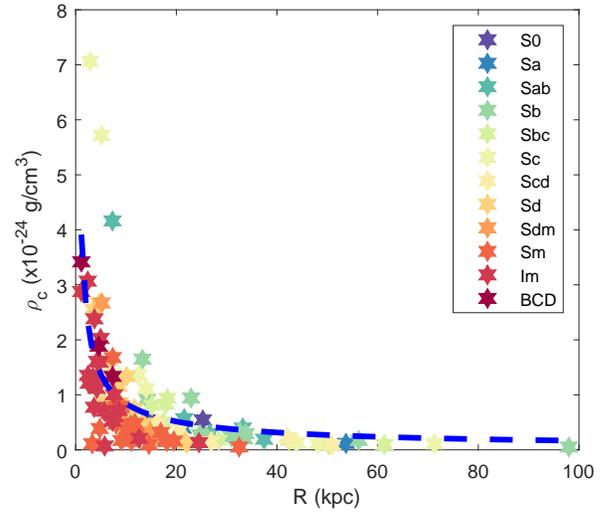}
\caption{\label{fig_R_ro}\small{The estimated values of the central density $\rho_c$ versus the estimated static radius of the dark matter halo $R$.
}}
\end{center}
\end{figure}

There is a power law type relation between the central density and the static radius of the halo $R$, given by
\be
 \frac{\rho_{c} (R)}{10^{-24}\;{\rm g/cm^3}}\approx  \alpha \times \left(\frac{R}{1\;{\rm kpc}}\right)^{-\beta},
\ee
where $\alpha=4.2732$, and $\beta =0.7060$, respectively. The correlation coefficient between  $\rho _c$ and $R$ is  $-0.4291$. Since the static radius is fully determined by the two basic parameters of the dark matter halo, the scattering length $a$ and the mass of the dark matter particle $m$, our analysis indicates the existence of a close correlation between these parameters and the central density of the dark matter halos.

\paragraph{The angular velocity distribution} The fitting results of the galactic rotation curves in the BEC model indicate a wide range of the numerical values of the angular velocity of the galactic halos. Fig.~\ref{fig_hist_omega} shows the distribution of the estimated values of the angular velocity $\omega$.

An interesting result is that for a large number of galaxies the angular velocity turns out to be very close to zero, indicating the static nature of the condensate dark matter halo. This raises the interesting question of the nature, origin and dynamics of the galactic rotation of dark matter, and of its dissipation for certain structures. Is the galactic rotation the result of some perturbations in the late evolutionary periods, or it is an intrinsic property that could be related to the initial collapsing phases? To answer these questions a detailed analysis of the formation of condensate structures in a cosmological setting is necessary. For rotating galaxies the angular velocity varied in the range of $\omega \approx 0.5\times 10^{-16}\; {\rm s}^{-1}$  and $\omega \approx 5\times 10^{-16}\; {\rm s}^{-1}$, respectively, indicating the presence of a gap in the rotational velocities. More exactly, it seems that there is no continuous transition or increase in the rotational velocity from zero to finite values, and the minimal rotation angular velocities having values of the order of $\omega \approx 0.5\times 10^{-16}\; {\rm s}^{-1}$.
\begin{figure}[h]
\begin{center}
\includegraphics[scale=0.6]{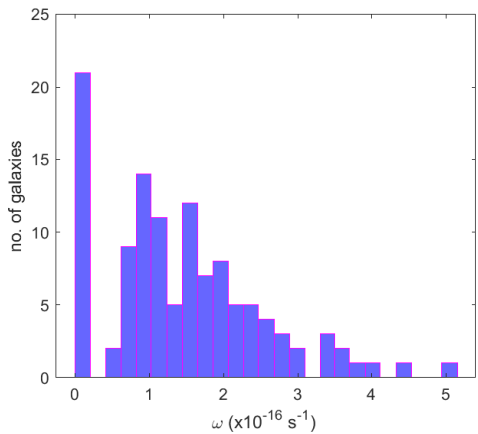}
\caption{\label{fig_hist_omega}\small{The distribution of the estimated values of the angular velocity $\omega$.
}}
\end{center}
\end{figure}

Fig.~\ref{fig_R_galtype_1} shows the values of the estimated angular velocity $\omega$ versus the estimated radius of the dark matter halo $R$  colored in function of galaxy type.
\begin{figure}[h]
\begin{center}
\includegraphics[width=\columnwidth]{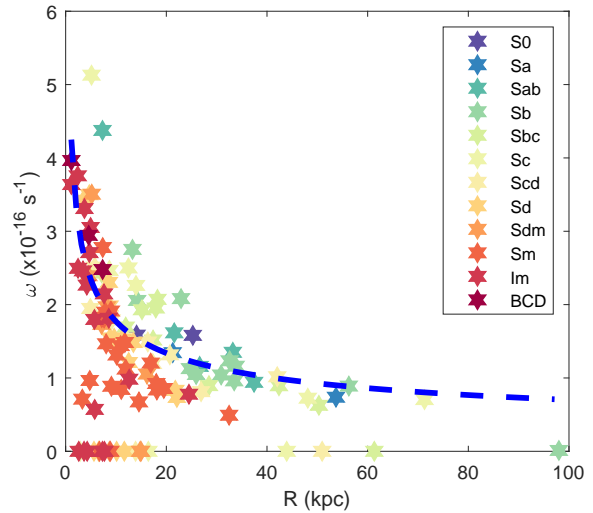}
\caption{\label{fig_R_galtype_1}\small{The estimated values of the  angular velocity $\omega$ versus the radius of the static dark matter halo $R$.
}}
\end{center}
\end{figure}

There is a power law relation between $\omega $ and $R$ of the form
\be
\frac{\omega (R)}{10^{-16}\;{\rm s}^{-1}} \approx \gamma \times\left( \frac{R}{1\;{\rm kpc}}\right)^{-\delta},
\ee
 where $\gamma =4.4686$ and $\delta=0.4022$. The correlation coefficient between $\omega $ and $R$ is  $-0.5854$.

The rotation of the BEC dark matter halos also raises the possibility of the existence of quantized vortices in this type of systems. The presence of vortices may bring new feature in the galactic dynamics and condensation, and could significantly affect dark matter properties \cite{inv40}.

\paragraph{The radius of the BEC halos.} In the present BEC model all the relevant properties of the dark matter halos are determined by three parameters only, the first two being the halo central density, and the rotation velocity of the galaxy. The third fundamental parameter, determining the structure and nature of the condensate dark matter halos is the radius $R$ of the static configuration.  $R$ is fully determined by the two fundamental physical parameters of the condensate dark matter, the mass $m$ of the dark matter particle, and by its scattering length $a$, respectively. Hence $R$ should have a universal value in all condensate dark matter models. In Fig.~\ref{fig_R_rmax_1} we have represented the distribution of the numerical values of $R$.
\begin{figure}[tbp]
\begin{center}
\includegraphics[scale=0.8]{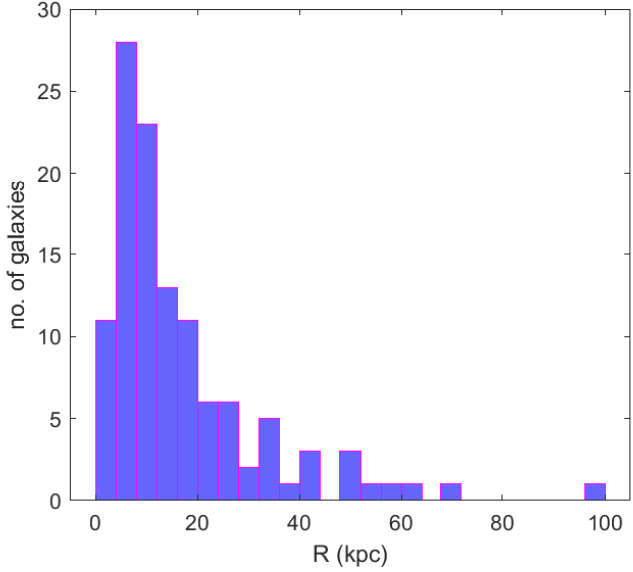}
\caption{\label{fig_R_rmax_1}\small{The distribution of the values of the estimated radius of the static dark matter halo $R$.
}}
\end{center}
\end{figure}

From our investigation it follows that $R$ is well constrained by the SPARC observational data, but the spread of the determined values of $R$ across the sample is very large, spanning roughly one order of magnitude. This finding seems to be incompatible with the existence of a single universal value for $R$. The BEC dark matter models considered in the present study make a very specific prediction about the existence of a finite, and well defined radius of the dark matter distribution. However, we must emphasize at this moment an important point. Obtaining the values of $R$ in our investigation is determined by the spatial extension of the analyzed sample.  Moreover, the values of $R$ are strongly correlated with the maximum value of $r_{max}$ from the observational data, indicating the known range of the extension of the rotation curves. In Fig.~\ref{fig_R_rmax} we have represented the correlation between the estimated values of the radius $R$ of the static dark matter configuration with respect to $r_{max}$, with empty markers for the galaxies for which we have obtained $\chi^2<5$, and with filled markers for the galaxies for which we have obtained $\chi^2<1$. These two quantities are strongly correlated, with the Pearson correlation coefficient given by $0.93$. The explicit relation between $R$ and $r_{max}$ is given by
\be
 R({\rm kpc})=0.9308\times r_{max}(\rm {kpc})+0.3584.
\ee

Hence in the fitting procedure and with the use of the present day data $R$ will always be located near the upper range of the radial extension of the observational data, and thus its numerical value will strongly depend on the quality and number of data, and of the distance from the galactic center where observations end.

\begin{figure}[h]
\begin{center}
\includegraphics[scale=0.8]{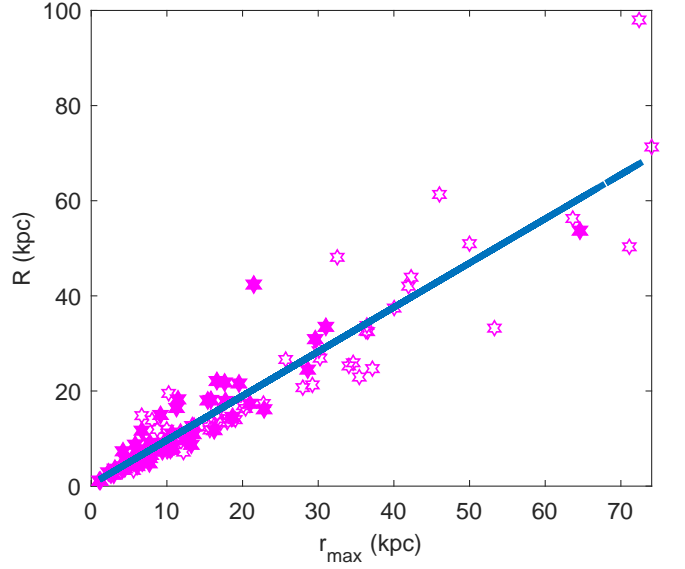}
\caption{\label{fig_R_rmax}\small{The estimated static radius of the condensate dark matter halo $R$ versus the maximum extension $r_{max}$ of the observational data.
}}
\end{center}
\end{figure}

In the case of rotating dark matter halos, the physical radius of the halo is not $R$, but the radius $R_h$, the value of the radial coordinate for which the density vanishes, and which defines the boundary radius of the dark matter distribution. $R_h$ depends on both angular velocity and central density of the halo. The distribution of the values of the halo radius $R_h$ are presented in Fig.~\ref{fig_R_h_1}.

\begin{figure}[tbp]
\begin{center}
\includegraphics[scale=0.8]{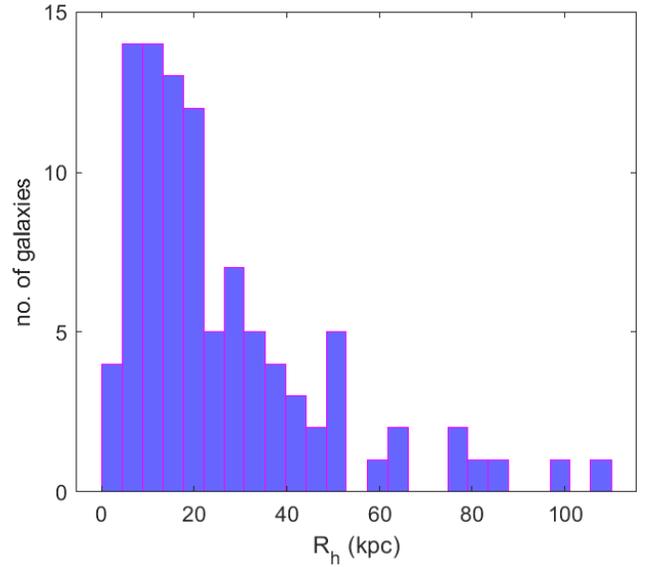}
\caption{\label{fig_R_h_1}\small{The distribution of the values of the galactic halo radius  $R_h$ for rotating galaxies with $\omega \neq 0$.
}}
\end{center}
\end{figure}

The radius of the rotating galactic halo shows more variability than the static radius $R$, whose distribution is peaked in the range of $R\in (0,10)$ kpc. One can also establish some power law type relations between the central density and the radius of the halo, and the angular velocity and the halo radius, given by
\be
\frac{\rho _c\left(R_h\right)}{10^{-24}\;{\rm g/cm^3}}\approx \alpha _h\left(\frac{R_h}{{\rm kpc}}\right)^{-\beta _h},
\ee
where $\alpha _h=5.1630$ and $\beta _h=0.6637$, with the Pearson correlation coefficient -0.4860, and
\be
\frac{\omega \left(R_h\right)}{10^{-16}\;{\rm s}^{-1}}\approx \gamma _h\left(\frac{R_h}{{\rm kpc}}\right)^{-\delta _h},
\ee
with $\gamma _h=5.2624$, and $\delta _h=0.3979$, respectively, with the Pearson correlation coefficient -0.6021.  There is also a very good correlation between $R_h$ and $r_{max}$, as given by
\be
R_h \left({\rm kpc}\right)=1.3058 \times r_{max} \left({\rm kpc}\right)+1.5163,
\ee
with the Pearson correlation coefficient given by 0.9442. This result clearly indicates that the estimation of the radius of the dark matter halo is strongly dependent on the number and quality of the observational data, since the objective function takes its minimal value on the boundary of the set of observations.

Presently there are no convincing data indicating the existence of physically well defined radii of the galactic dark matter halos, and, for non-rotating galaxies,  of their universal nature. In several dark matter models the dark matter distribution extends to infinity. However, if a future detection of the radii of a large sample of galaxies would be possible, this would allow a definite test of the BEC dark matter model. An observational proof of the  infinite extension of the galactic halos would certainly contradict one of the basic predictions of the BEC model.

\paragraph{The total mass distribution of the dark matter halo} The total mass of the galaxy is an important observational parameter that can be used to test and distinguish between different dark matter models. The mass of a galaxy and its distribution can be generally  obtained by using two methods. In the first method one can use photometric data, such as luminosity profiles,  which implies some assumptions on the mass-to-luminosity ratio.  The second approach consists in applying dynamical methods to the kinematical data by using the virial theorem. Obtaining the total mass of the galaxy, including the dark matter and invisible masses like black holes, requires the application of the dynamical methods, with the total mass calculated under the assumption that the kinetic and gravitational energies are in equilibrium for a relaxed astrophysical system. Another important method for the mass determination is based on the rotation curves \cite{New1}. In order to obtain the total galactic mass a constant mass-to-light ratio is usually attributed to the luminous disk, and the parameters of the dark matter halo are obtained via the best fit to the observed rotation curve. There are still some problems with this method, since the numerical values of the mass-to-light ratio for the luminous matter are not known precisely. Moreover, in many cases the  functional form for density or mass distributions of the dark matter halos are adopted arbitrarily.

On the other hand in the Bose-Einstein Condensate dark matter model there is an exact predictions of the total mass $M$ of the dark matter halo, which follows immediately from  Eq.~(\ref{masseqn}), giving for $M$ the expression \cite{Zhang}
\be
M_{DM}\left(R, \rho _c,\Omega\right)=\frac{4}{\pi ^2}\rho _cR^3\left[1+\left(\frac{\pi ^2}{3}-1\right)\Omega ^2\right].
\ee

The total mass of the Bose-Einstein Condensate halo is fully determined by three basic parameters: the static radius of the halo, its central velocity, and the rotational angular velocity, respectively. Once this parameters are known, we can make a consistent and precise prediction of the mass of the dark component of the galaxy. The total masses of a selected set of galaxies from the SPARC sample are presented in Table~\ref{table:MyTableLabel}.

\begin{widetext}
\begin{center}
\begin{table}
\begin{footnotesize}
\centering
\begin{tabular}{|l|r||l|r||l|r||l|r|}\hline

Galaxy & $\!\!$ Mass \ \ \ \    & Galaxy & $\!\!$  Mass \ \ \ \      & Galaxy & $\!\!$  Mass\ \ \ \   & Galaxy & $\!\!$  Mass \ \ \ \   \\

\     &   $\left(10^{10} M_{\odot}\right)$ & \ &  $\left(10^{10} M_{\odot}\right)$& \ &  $\left(10^{10} M_{\odot}\right)$ & \ &  $\left(10^{10} M_{\odot}\right)$\\ \hline
DDO064       & 0.099 & NGC0055      & 1.378 & NGC4100      & 6.134 & UGC06917     & 1.353 \\
 \hline
DDO161       & 0.988 & NGC0100      & 1.237 & NGC4214      & 0.577 & UGC06930     & 2.959 \\
 \hline
DDO168       & 0.236 & NGC0289      & 25.487 & NGC4559      & 3.780 & UGC06983     & 1.993 \\
 \hline
ESO079-G014  & 6.616 & NGC0300      & 1.687 & NGC7793      & 0.855 & UGC07089     & 1.556 \\
 \hline
ESO116-G012  & 1.706 & NGC1705      & 0.423 & UGC01230     & 4.207 & UGC07125     & 0.712 \\
 \hline
ESO444-G084  & 0.290 & NGC2366      & 0.194 & UGC01281     & 0.271 & UGC07151     & 0.229 \\
 \hline
F563-1       & 3.556 & NGC2915      & 1.237 & UGC04278     & 2.214 & UGC07261     & 1.969 \\
 \hline
F563-V2      & 1.037 & NGC2976      & 0.317 & UGC04483     & 0.010 & UGC07323     & 0.799 \\
 \hline
F565-V2      & 1.053 & NGC2998      & 21.486 & UGC04499     & 0.553 & UGC07524     & 0.570 \\
 \hline
F568-1       & 2.032 & NGC3109      & 0.623 & UGC05005     & 4.529 & UGC07559     & 0.040 \\
 \hline
F568-3       & 2.727 & NGC3521      & 11.144 & UGC05716     & 0.912 & UGC07603     & 0.264 \\
 \hline
F568-V1      & 3.638 & NGC3726      & 29.701 & UGC05721     & 0.649 & UGC07608     & 0.390 \\
 \hline
F571-8       & 6.168 & NGC3741      & 0.308 & UGC05750     & 2.376 & UGC07690     & 0.126 \\
 \hline
F571-V1      & 1.465 & NGC3893      & 7.095 & UGC05918     & 0.032 & UGC08286     & 1.764 \\
 \hline
F574-1       & 1.082 & NGC3972      & 2.479 & UGC05986     & 1.566 & UGC08490     & 0.896 \\
 \hline
F583-1       & 1.818 & NGC3992      & 40.351 & UGC06399     & 0.752 & UGC08550     & 0.289 \\
 \hline
IC2574       & 2.768 & NGC4010      & 2.589 & UGC06446     & 0.829 & UGC08837     & 0.382 \\
 \hline
KK98-251     & 0.071 & NGC4051      & 1.104 & UGC06628     & 0.090 & UGC09037     & 7.818 \\
 \hline
NGC0024      & 3.389 & NGC4085      & 1.816 & UGC06667     & 0.622 & UGC10310     & 0.301 \\
 \hline
NGC0055      & 1.378 & NGC4088      & 30.925 & UGC06818     & 0.990 & UGC11557     & 1.086 \\
 \hline

\end{tabular}
\caption{The predicted masses of the dark matter halos for a selected set of galaxies of the SPARc sample.}
\label{table:MyTableLabel}
\end{footnotesize}
\end{table}
\end{center}
\end{widetext}

The predicted masses of the dark matter halos lie in the range of $10^{10}-10^{11}M_{\odot}$, which is consistent to our present knowledge of galactic properties. The mass distribution of the galaxies as a function of the number of galaxies in the sample is presented in Fig.~\ref{fig_hist_mass}.

\begin{figure}[tbp]
\begin{center}
\includegraphics[scale=0.7]{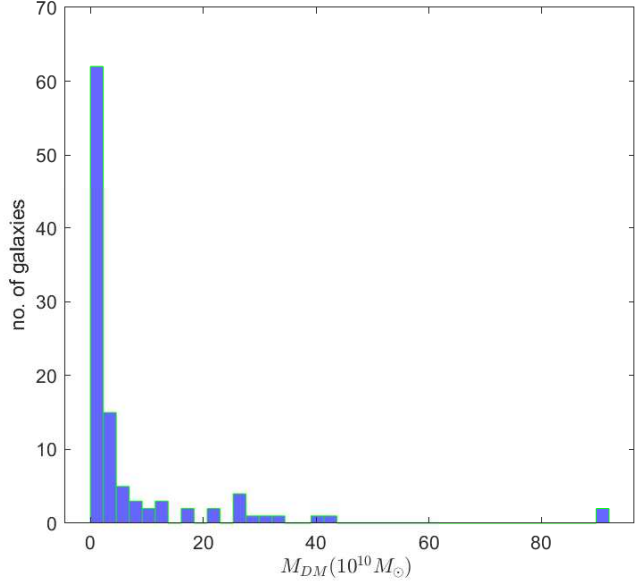}
\caption{\label{fig_hist_mass}\small{The distribution of the galactic masses of the Bose-Einstein Condensate dark matter halos.
}}
\end{center}
\end{figure}

\paragraph{The mass and scattering length of the dark matter particle.} The mass $m$ of the dark matter particle, and the scattering length $a$, describing the self-interacting properties of the condensate, are the basic parameters describing the physical properties of galactic halos.  Unfortunately in the present theoretical approach these parameters cannot be constrained independently, since only their combination $a/m^3$ does appear in the expression of the static radius of the condensate, giving
\be
\frac{a}{m^3}=\frac{G}{\hbar ^2}\left(\frac{R}{\pi}\right)^2=5.77\times 10^{90}\times \left(\frac{R}{10\;{\rm kpc}}\right)^2\;\frac{{\rm cm}}{{\rm g^3}}.
\ee

The distribution of the values of $a/m^3$ as obtained from the SPARC sample is presented in Fig.~\ref{am3} . There is a sharp maximum of this ratio, for around $80$ galaxies this ratio shows a roughly constant value,  with the upper limit of the first bin including the values for 75 galaxies being $a/m^3=0.1390\times 10^{92}\;{\rm cm/g^3}$. On the other hand the minimum of the values of $a/m^3$ obtained from our analysis of the SPARC sample is $a/m^3=7.4032\times 10^{88}\;{\rm g/cm^3}$.  Hence our present results show that the possibility that at galactic level $a/m^3$ is an absolute constant cannot be excluded a priori, and in fact the observational evidence tends to point into this direction. However, the present day observational data cannot provide a definite answer to this question.

\begin{figure}[tbp]
\begin{center}
\includegraphics[scale=0.7]{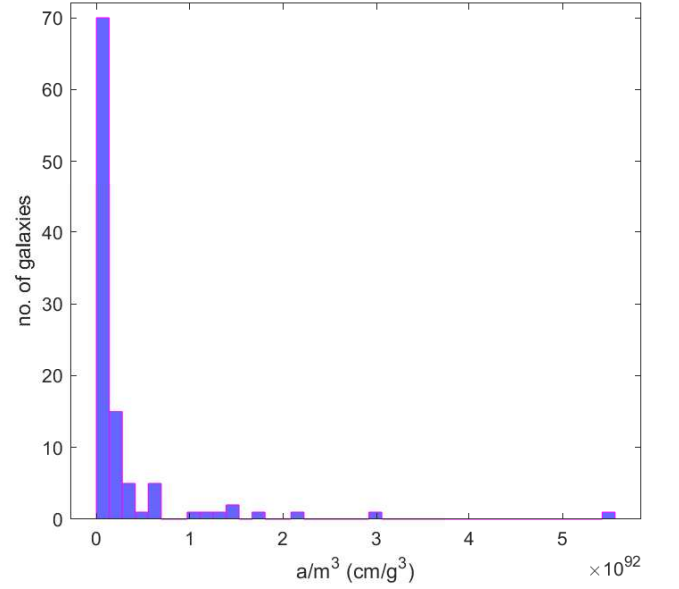}
\caption{\label{am3}\small{The distribution of the values of the estimated ratio of the dark matter scattering length and mass $a/m^3$.
}}
\end{center}
\end{figure}

Some theoretical constraints on the ratio $a/m^3$ can be obtained from the condition $2GM_{J}/c^{2}R_{J}<1$, where $M_J$ and $R_J$ are the Jeans mass and the Jeans radius of the condensate. This condition, which is essentially general relativistic, requires that dark matter halos cannot become black holes.
By assuming  that it is also valid
in Newtonian mechanics, and by  adopting for the Jeans mass and radius  the expressions \cite{inv40},
\begin{eqnarray}  \label{34}
M_J&\approx &\frac{32\pi^4}{3}\left(\frac{\hbar ^2a}{Gm^3}\right)^{3/2}\rho
_0=2.623\times 10^{12} \times  \notag \\
&&\left(\frac{a}{10^{-3}\;\mathrm{fm}}\right)^{3/2} \times\left(\frac{m}{%
\mathrm{meV}}\right)^{-9/2}\times  \notag \\
&&\left(\frac{\rho _0}{10^{-24}\;\mathrm{g/cm^3}}\right)M_{\odot},
\end{eqnarray}
and
\begin{eqnarray}\label{Jeans1}
\hspace{-0.5cm}R_J&\approx& \sqrt{\frac{\pi}{G\rho _0}}v_s=\sqrt{\frac{4\pi
^2\hbar ^2a}{Gm^3}}  \notag \\
\hspace{-0.5cm}&& =34.96\times\left(\frac{m}{\mathrm{meV}}\right)^{-3/2}\left(\frac{a}{%
10^{-3}\;\mathrm{fm}}\right)^{1/2}\;\mathrm{kpc},
\end{eqnarray}
respectively, where $\rho _0$ is a specific density, and $v_s$ is the speed of sound in the condensate, we obtain the constraint
\begin{equation}
\frac{a}{m^3}<\frac{3c^2}{32\pi ^3\hbar ^2\rho _0}=2.449\times 10^{96}
\left( \frac{\rho _{0}}{10^{-24}\;\mathrm{g/cm^{3}}}\right) ^{-1}\;\mathrm{%
\frac{cm}{g^3}}.
\end{equation}

The observational results from the SPARC sample indicate that this upper bound for $a/m^3$ is certainly satisfied.

\subsection{Correlations with the galactic properties}

As we have already mentioned, a detailed analysis of a version of the Bose-Einstein Condensate dark matter models and the galactic properties of the SPARC sample was performed in \cite{New}, for a model that is distributed in several states, and at a nonzero temperature. In the following we will investigate this problem in the framework of the present zero temperature BEC theory.
\paragraph{Static radius-galactic distance correlation} The study of the correlation between the static radius $R$ and the distance to the galaxy is one of the most important correlations for the understanding of the dark matter halo properties. The presence of a null correlation implies that the sizes of the dark matter  halos do not depend on the distance to the host galaxy \cite{New}. The correlation between $R$ and the distances to the galaxies is presented in Fig.~\ref{fig_R_Distance}.

\begin{figure}[tbp]
\begin{center}
\includegraphics[scale=0.8]{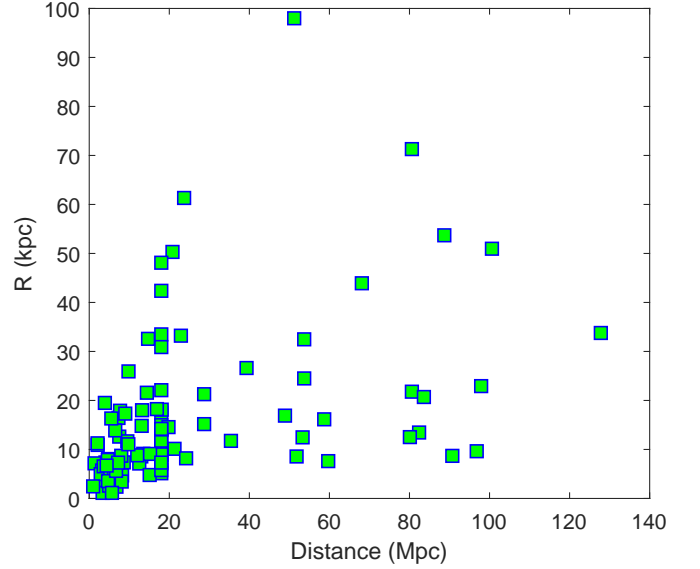}
\caption{\label{fig_R_Distance}\small{The values of the static radius versus the galactic distances.
}}
\end{center}
\end{figure}

The Pearson correlation coefficient between the galactic distances and the static radius of the dark matter halo $R$ is 0.4503, indicating the possibility of the existence of a weak correlation between these two quantities, a result similar to the one obtained in \cite{New}. In fact, in \cite{New}, it was shown that after removing galaxies located at distances greater than 80 Mpc, the correlation coefficient becomes 0.22, a result which is consistent with the null hypothesis.

\paragraph{Dark halo mass and distance correlation.} The correlation between the dark matter halo mass and the galactic distance is presented in Fig.~\ref{fig_MDM_distance}. The correlation coefficient is 0.3998, still indicating the possibility of the existence of a weak correlation between the mass of the dark halo, and the galactic distances.

\begin{figure}[tbp]
\begin{center}
\includegraphics[scale=0.8]{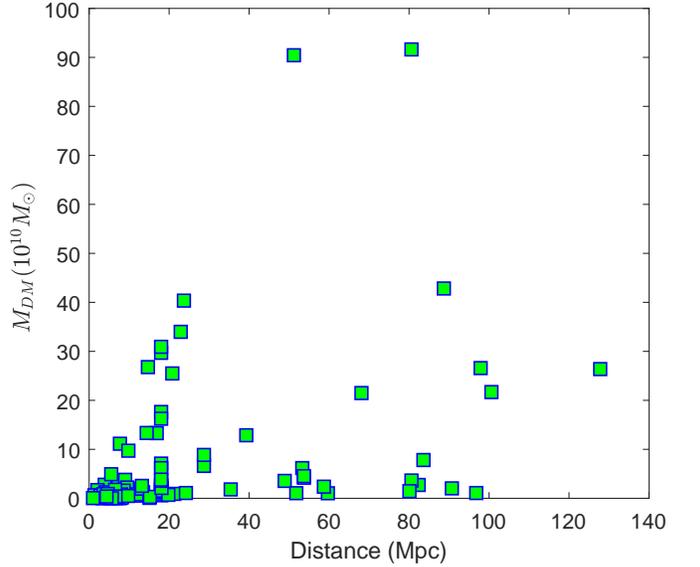}
\caption{\label{fig_MDM_distance}\small{The values of the mass of the dark matter halo versus the galactic distances.
}}
\end{center}
\end{figure}

\paragraph{Correlation between galactic luminosity and the static radius of the dark matter halo.} We have found an acceptable correlation between the static radius $R$ of the dark matter halo, and the total luminosity of the baryonic matter in the galaxy. The correlation between these two parameters is represented in Fig.~\ref{fig_R_Total_Lum}. The correlation coefficient is 0.7110, indicating the possibility of the existence of a significant relation between the dark matter and baryonic matter properties, which could be a consequence of the interplay between ordinary and dark matter during the epoch of structure formation in the early Universe.

\begin{figure}[tbp]
\begin{center}
\includegraphics[scale=0.8]{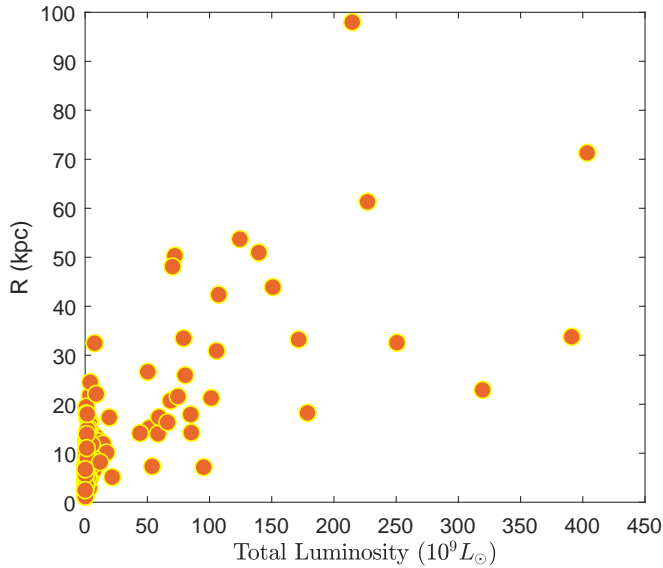}
\caption{\label{fig_R_Total_Lum}\small{The values of the static radius of the dark matter halo versus the total luminosity.
}}
\end{center}
\end{figure}

\paragraph{Static radius versus total $M_{HI}$ mass.} We have found a good correlation between the total HI Mass of the galaxy $M_{HI}$, and  the  static radius of the dark matter hallo. The correlation is presented in Fig.~\ref{fig_R_MHI}. The correlation coefficient is 0.8530, and the linear relationship between these two quantities is given by the equation
\be
R ({\rm kpc})\approx 1.0022\times M_{HI}\left(10^9M_{\odot}\right)+ 0.3168.
\ee

\begin{figure}[tbp]
\begin{center}
\includegraphics[scale=0.8]{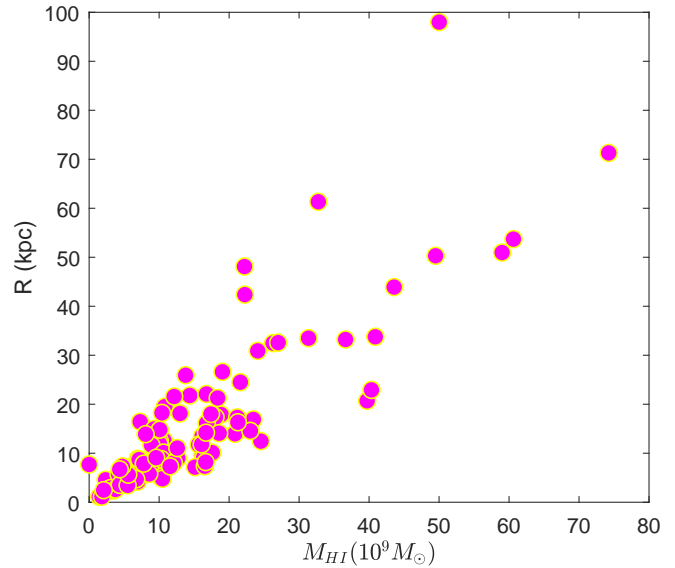}
\caption{\label{fig_R_MHI}\small{The values of the static radius of the dark matter halo versus the total HI mass of the galaxy.
}}
\end{center}
\end{figure}

This correlation can be understood in the sense that an increase of the galactic radius implies the existence of more baryonic mass, and, consequently, of more neutral hydrogen in the galaxy.

\paragraph{Total luminosity and the mass of dark matter halo correlation.} There is also a good correlation between the total mass of the dark matter halo and the total luminosity of the galaxy, which is represented in Fig.~\ref{fig_MDM_Lum}. The correlation coefficient between  the total luminosity and the total mass of dark matter $M$ is 0.8085, and the relation between these two quantities can be represented by the linear equation
\be
M_{DM}\left(10^{10}M_{\odot}\right)\approx 0.1539\times L\left(10^9L_{\odot}\right)   + 1.0135.
\ee

\begin{figure}[tbp]
\begin{center}
\includegraphics[scale=0.8]{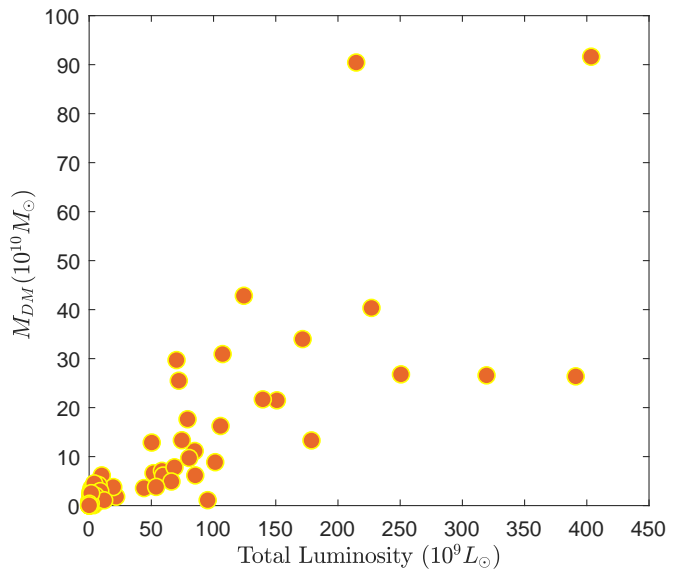}
\caption{\label{fig_MDM_Lum}\small{The values of the mass of the dark matter halo versus the total luminosity of the galaxy.
}}
\end{center}
\end{figure}

The existence of such a relation indicates again that the more massive a galaxy is, the more luminous baryonic matter it contains. These correlations are a result of the complex cosmological and astrophysical processes taking place during the galaxy formation stage in the early Universe.

\paragraph{Other correlations.} We would also like to briefly mention a number of other correlations between dark matter and galactic properties we have investigated. We have found a good correlation between the asymptotically flat rotation velocity $v_{flat}$ and the static radius of dark matter halo $R$, with the correlation coefficient given by 0.7401. Similarly, there is a trend indicating the existence of a correlation between the Effective Radius at [3.6] of the galaxy,  and the  static radius of the dark matter $R$. The correlation coefficient between these two quantities with data extracted from the SPARC sample is 0.6596.

There is a poor correlation between the central dark matter density, and the galactic properties. The correlation coefficient between  galactic distance and the central density of the dark matter  $\rho_c$ is  -0.2517, a value that  is consistent with  the null correlation between these two quantities. The correlation  between the total luminosity of the galaxy and the central density of the condensate dark matter  $\rho_c$ is -0.1964, pointing towards the nonexistence of any correlation between these two quantities. We would also like to point out that the central density of the dark matter halo is anti correlated with both galactic distances and the baryonic luminosity of the galaxy.

\section{Discussions and final remarks}\label{sect3}

A large number of cosmological and astrophysical observations, including the intriguing behavior of the galactic rotation curves, as well as the virial mass discrepancy in galaxy clusters, strongly point towards the necessity of the existence of a fundamental component of the Universe, conventionally called  dark matter. Usually it is assumed that dark matter is cold, and pressureless. If dark matter is present as a gas of bosonic particles, then the laws of quantum mechanics requires that it is in the form of a Bose-Einstein Condensate. According to standard statistical physics, in a boson gas a phase transition to a condensate state must automatically occur, once the temperature reaches its critical value. Since the present day temperature of the Universe is lower than the critical transition temperature, it is legitimate to assume that dark matter stands as a Bose-Einstein Condensate. If observations would confirm this important (and intriguing) hypothesis,  major modifications and adjustments of our interpretations of the basic principles of cosmology and  astrophysics would be needed. In the present study we have presented some basic results about the observational significance and interpretation of the Bose-Einstein Condensate dark matter model, and of its possible astrophysical relevance. These results could help in discriminating this model from the plethora of already proposed alternative dark matter models. Moreover, our results also provide some strong constraints on the free parameters of the model.

In the present work, our main focus was the detailed investigation of the potential observational signatures of the Bose-Einstein Condensate dark matter model in the high-quality HI/H$\alpha$ measurements of the rotation curves of the SPARC data. As a first step in our analysis we did carefully reconsider the data of the SPARC sample, by excluding all galaxies with a low number of data points. Hence in our study we have considered only 157 of the original 175 galaxies. We have modeled the galaxies in the SPARC database as containing a stellar disk, a gaseous disk, a stellar bulge, if necessary, and the condensate dark matter halo, as the main component essentially determining the behavior of the galactic rotation curves. After selecting our sample, we have fitted the observational data on the rotational velocities by using  Eq.~(\ref{eq_vh}), which  gives the expression of the tangential velocity of massive test particles in the slowly rotating  Bose-Einstein Condensate dark matter model. Note that the Bose-Einstein Condensate dark matter models predicts a tangential velocity of massive particle that is determined by three, astrophysically observable, quantities only:  the radius $R$ of the static configuration,  the angular velocity of the rotating dark matter distribution $\omega $, and by the central density $\rho _c$ of the halo, respectively.

Our results show that the tangential velocity expression of the Bose-Einstein Condensate dark matter model gives a good fit of the SPARC observational data, with a value of $\chi ^2$ smaller than 5 for a number of 117 galaxies, from the total of the 157 considered. Hence for around 75\% of the representative galaxies of the SPARC sample the fittings of the rotation curves with the model can be considered very good/good. For the rest of the galaxies the fits are also acceptable, as judged according to the value of $\chi ^2$. But in this case we must mention that the value of $\chi ^2$ may exceed 5. A good fit also requires numerical values of the fitting parameters beyond our considered (and rather strict) range. The fitting procedure provides us with some intrinsic properties of the condensate dark matter halo, like its central density, rotational velocity, and static radius, and with the stellar mass to light ratios for the disk and the bulge, respectively. As expected from astrophysical and cosmological considerations, there is a large range in the distribution of the central densities and angular velocities, which may be attributed to the different environmental conditions and gravitational interactions in which galaxy formation took place. The density distribution of the galactic dark matter halo can be completely reconstructed, and the density profile can be exactly predicted. As one can see from Fig.~\ref{fig_density}, a characteristic and significant property of the Bose-Einstein Condensate dark matter halos is the presence of a well-defined radius (boundary) $R_h$ of the galaxy, on which the dark matter density vanishes. Condensate dark matter cannot extend beyond $R_h$. Most of the alternative dark matter profiles allow an infinite extension of dark matter, without the presence of an exactly defined radius, like, for example, the Navarro-Frenk-White density profile \cite{cusp}. The radius of the dark matter $R_h$ depends on the rotation velocity $\omega $ of the galaxy, and on an universal parameter, the radius of the static condensate configuration $R$. Due to the rotation effects the radius of the galaxy $R_h$ shows a large degree of variability. On the other hand a large degree of variability of $R$ does appear in the static (nonrotating galaxies). If we eliminate a number of pathological cases, like NGC2998 ($R=43.916$ kpc), NGC3992 ($R=61.324$ kpc) and UGC12506 ($R=50.963$ kpc), for static galaxies $R$ takes values in the range of 3 to 16 kpc, with an average of around 10 kpc. Precise observational determinations of the galactic radii will provide a definite test of the Bose-Einstein Condensate dark matter model.

An important and attractive property of the density distribution profile of the Bose-Einstein Condensate dark matter is that it is naturally nonsingular at the galactic center. Moreover, it has a clearly defined cored central profile. It has already been suggested that the Bose-Einstein condensation of dark matter can straightforwardly solve the core-cusp
problem \cite{Har1}, and hence eliminate the unphysical and less attractive features that characterize the Navarro-Frenk-White, $\Lambda$CDM type, dark matter density profile, obtained from extended numerical simulations \cite{cusp}. Our results and theoretical predictions of the dark matter density distribution, based on the SPARC
sample, positively indicate the presence of a dark matter core at the galactic center, a result which is also backed by many other astrophysical observations.

We have also considered in detail the relations between the defining parameters of the condensate dark matter (static radius and total mass) and the galactic properties. We have found that a significant degree of correlation does exist between the static radius of the dark matter halo $R$ and the luminosity of the galaxy, the asymptotically flat rotation velocity $v_{flat}$, the Effective Radius at [3.6] and $R$,  and  the total HI Mass, respectively. There is also a good  correlation between the total luminosity and the total mass of the dark matter halo $M$. We have also detected a possible trend indicating the existence of a weak correlation between galactic distance and the static radius of the dark matter halo, and  between distance and the total mass of the condensate dark matter halo. We have found a null correlation between the numerical values of the central dark matter density, and the galactic properties.

It would be extremely interesting and important to compare the present model, essentially based on the Bose-Einstein Condensate density profile, with other dark matter models, like, for example, the Navarro-Frenk-White dark matter density distribution \cite{cusp}. One such possibility of comparison would be the use of the Akaike Information Criterion (AIC) \cite{new2} or the  Bayesian Information Criterion (BIC) \cite{new3}. The AIC and BIC criteria are used as model selection criteria, that is, their main goal is to compare the fitting of different models on the same data, taking into account the number of parameters to be optimized. They penalize models with a larger number of parameters. However, they are not relevant when we are interested in the fitting of a single model, like in the present investigation. In our case the quality of the fitting is evaluated by using the numerical values of $\chi^2$.

Unfortunately the Bose-Einstein Condensate dark matter model in its present formulation cannot provide any firm prediction on the mass of the dark matter particle. All the properties of the galaxy depend on the combination of two basic physical parameters, the dark matter particle scattering length, and the mass of the dark matter particles, which appear in the combination $a/m^3$. Assuming a self-interaction cross section of the dark matter particles of the order of $\sigma _m=1.25\;{\rm cm^2/g}$, and for the static radius $R$ a mean value of the order of 10 kpc, the mass of the dark matter particles, as given by Eq.~(\ref{massdm}), should be of the order of $m=0.2$ meV. Different mass values can be obtained easily by slight modifications of the numerical values of $\sigma _m$ and $R$. Such a mass range does not seem to correspond to any known particle, or to the standard model of the particle physics. The closest candidate for a particle with such a mass may be the neutrinos, for which the analysis of the data from  the Baryon Oscillation Spectroscopic Survey, from the Sloan Digital Sky Survey, and from the Planck 2015 Cosmic Microwave Background observations gives the constraint $\sum{m_{\nu}} < 0.12$ eV (95\% C.L.) \cite{neutr}. The most stringent present constraint on the effective neutrino mass $m_{\beta}$ is $m_{\beta}<1.1$ eV, with a lower bound of the order $m_{\beta}>50$ meV \cite{neutr1}. In \cite{neutr2} it has been suggested that the massive neutrino states are in fact Bogoliubov quasiparticles,  and their vacuum is a condensate of "Cooper pairs" of massless flavor neutrinos. Note that from a physical point of view the neutrino condensation may arise due to the presence of some attractive interactions between the particles. In the case of the Dirac neutrinos,  the interaction with the Higgs field may lead to their condensation. The implications of neutrino self-interactions for the physics of weak decoupling and Big Bang Nucleosynthesis  in the early Universe were recently considered in \cite{Grohs}. The effects of neutrino self-interaction for the primordial helium, deuterium abundances for a measure of relativistic energy density at photon decoupling were investigated. Self-interacting neutrinos can also possibly solve a number of cosmological anomalies, including the amelioration of the tension in the Hubble parameter.  Constraints on the masses and properties of light sterile neutrinos coupled to all three active neutrinos were obtained in \cite{Sunny}, by using a combination of cosmological data and terrestrial measurements from oscillations, $\beta$-decay and neutrinoless double-$\beta$ decay experiments.  When allowing for the  mixing with a light sterile neutrino, from cosmological considerations one arrives at upper bounds for the electron neutrino mass $m_{\beta}$ and the Majorana mass parameter $m_{\beta \beta}$, as given by
$m_{\beta} < 0.09$ eV and $m_{\beta \beta} < 0.07$ eV at 95\% C.L.

Alternative dark matter candidates may be represented by the axions, hypothetical particles with masses of the order of $10^{-22}$ meV, having a de Broglie wavelength $\lambda \sim 1 $ kpc \cite{Hui}. Note that the axions based dark matter models are called fuzzy dark matter models. Fuzzy dark matter halos consist of a core represented by a stationary, minimum-energy configuration, also called a "soliton" \cite{Hui}. The core is surrounded by an envelope that resembles a cold dark matter halo. Note that the axions can condensate, and form a Bose-Einstein Condensate, since their mass obviously satisfy the constraint (\ref{massdm}). But in the framework of the present condensate dark matter models it is impossible to make a convincing prediction on the numerical value of the mass of the dark matter particle. However, advances in the understanding of the self-interaction and condensation processes of low mass elementary particles may lead to a precise theoretical estimation of the scattering length, and this would open the possibility of the precise prediction of the particle mass of the condensate dark matter halos.

The final proof of the possible existence of the Bose-Einstein Condensate dark matter in nature substantially depends  on the significant increase in the number of observational data points of the galactic rotation curves, and on their precision. Measurements of the rotation curves that extend far from the galactic center could answer to the question if dark matter is strictly confined within a finite radius, or if it extends to infinity, as assumed in most standard dark matter models. The existence of a universal length scale, the galactic radius, and fully determined by the physical properties of the dark matter particle, is one of the main predictions of the Bose-Einstein Condensate dark matter theory. On the other hand, the final outcome of the core/cusp controversy may also provide evidence for the existence of Bose-Einstein Condensate dark matter.  If extensive numerical simulations done in the framework of the standard $\Lambda$CDM cosmological model, with all baryonic effects included, could predict cored central density galactic profiles, then the presence of galactic halos made of Bose-Einstein Condensate dark matter may not be necessary.  If, on the other hand, the results of the realistic numerical simulations would still show the presence of cuspy Navarro-Frenk-White type galactic halo profiles, then the Bose-Einstein Condensate dark matter model may provide a realistic  and physical explanation for the cored central structures observed in a large number of galaxies.

\section*{Acknowledgments}

We would like to thank to the anonymous referee for comments and recommendations that helped us to improve our manuscript. We also thank Prof. L. \'{A}. Gergely for useful suggestions and help.

\end{document}